\documentclass[11pt]{article}
\usepackage{graphicx}
\usepackage{amsmath}
\begin{document}

\begin{center} 
{\Large\bf   QCD Factorization of Quasi Generalized Gluon Distributions  }
\par\vskip20pt
 J.P. Ma$^{1,2,3}$,  Z.Y. Pang$^{1,2}$, C.P. Zhang$^{1,2}$ and G.P. Zhang$^{4}$    \\
{\small {\it
$^1$ CAS Key Laboratory of Theoretical Physics, Institute of Theoretical Physics, P.O. Box 2735, Chinese Academy of Sciences, Beijing 100190, China\\
$^2$ School of Physical Sciences, University of Chinese Academy of Sciences, Beijing 100049, China\\
$^3$ School of Physics and Center for High-Energy Physics, Peking University, Beijing 100871, China\\
$^4$ Department of Physics, Yunnan University, Kunming, Yunnan 650091, China}} \\
\end{center}

\vskip 1cm
\begin{abstract}
We study the factorization relations between quasi gluon GPDs and twist-2 GPDs.
The perturbative coefficient functions are obtained at one-loop level. They are free from any collinear- or I.R.  divergences. Unlike the case of the factorization of quasi quark GPDs at one-loop, we have to add ghost 
contributions for the factorization of quasi gluon GPDs in order to obtain gauge-invariant results.
In general, operators will be mixed beyond tree-level. Our work shows that the mixing pattern 
of the nonlocal operators in quasi gluon GPDs is the same as those of local operators, i.e., the nonlocal operators considered are mixed with gauge-invariant operators, BRST-variation operators and operators involving EOM operator.  The factorization relations are obtained for all quasi gluon GPDs. 
Taking the forward limit, we also obtain the relations between quasi gluon PDFs and twist-2 PDFs.

\end{abstract}      
\vskip 5mm

\vskip40pt

\noindent
{\bf 1. Introduction} 

\par\vskip5pt

Inside a hadron there are not only quarks and antiquarks but also gluons.  It is important to understand the role played by these gluons for explaining properties of hadrons.  In high-energy scattering, the effects of the gluons can be characterized by various gluon distributions of hadrons based on QCD factorization theorems. 
A well-known example is twist-2 gluon distribution functions. They and   twist-2 quark distribution functions are called as Parton Distribution Functions(PDFs)  and used extensively for making predictions 
of inclusive processes. Generalizing twist-2 parton distributions one obtains Generalized Parton Distributions(GPDs).     

GPDs have been introduced in \cite{DMGPD,DVCSJi}.  These distributions contain more information about hadron's inner structure than PDFs. e.g., they contain information about contributions from quarks and gluons to the spin of a proton as shown in \cite{DVCSJi}.  Because of their importance,  extensive study of GPDs has been performed and its results can be found in reviews \cite{DMGPDR,MDI,BeRa}. 
Since GPDs and PDFs represent long-distance effects of QCD, they can not be predicted with perturbative QCD. Predictions of nonperturbative effects can be made by Lattice QCD from the first principle. However, a direct calculation of PDFs and GPDs as distributions is not possible, because  
they depend on time in Minkowski space. Only moments of these distributions can be calculated with Lattice QCD. 

A new method called as  Large-Momentum Effective Field Theory(LaMET) in \cite{Ji,JiC} has been proposed to calculate PDFs directly.  In this method one defines quasi PDFs
with time-independent operators. Since  quasi PDFs have no time-dependence, they 
can be directly calculated with Lattice QCD.  
The defined quasi PDFs are related to PDFs 
in the limit of large hadron momentum. 
The relation 
is perturbative, or the quasi PDFs can be factorized with twist-2 PDFs in the limit. 
This proposal has stimulated intensive studies of calculations of parton distributions,  a review about current progresses in this field can be found in \cite{LAM}. The proposed method can be used for GPDs. 
In this work we study the factorization of gluon quasi GPDs. The factorization of quark quasi GPDs has been studied in \cite{MGPD1, MGPD2, MGPD3,MaPaZh}. 

In this work we employ the same approach as that used for the factorization of quark quasi GPDs in \cite{MaPaZh}. The approach is based on diagram 
expansion which has been successfully used for the analysis of higher-twist effects in DIS in \cite{EFP, JWQ}. With the approach one is able to directly calculate perturbative coefficient functions in the factorization. There is no need to calculate quasi GPDs and GPDs of parton states.  
The factorization relations derived in this work  are between matrix elements used to define quasi gluon GPDs and those used to define twist-2 GPDs. Individual quasi gluon GPDs defined with one operator has the same factorization relation to the corresponding twist-2 GPDs. The obtained relations apply to the cases of hadrons with any spin number.

In deriving the factorization relation of quasi quark GPDs, gauge invariance is relatively easy to maintain
at one-loop. 
In the case of quasi gluon GPDs studied here, it becomes more complicated. At first look, there are contributions of super-leading-power in Feynman gauge. The existence of such contributions in factorizations has been first pointed out in \cite{JCTR}. Such contributions are obviously gauge variant.   
With a Ward identity we are able to show that all super-leading power contributions are in fact zero. 
However, at the leading power one still can not obtain gauge-invariant results directly. One needs to consider 
contributions from ghost fields. 
Including these contributions 
and using Equation Of Motion(EOM) of QCD we are able to derive gauge-invariant results.  At operator-level, 
our results indicate that the gauge-invariant operators used to defined quasi gluon GPDs are only mixed with
those operators: Gauge-invariant operators, BRST-variation operators and operators involving the EOM operator. This mixing pattern may be expected. Indeed there is a proof for such a mixing pattern but only for local operators in \cite{JoLee,Jo}. It is noted that operators of quasi GPDs can not be represented  
 by local operators which are related to moments of quasi GPDs. This is because 
 that the moments can not be defined for quasi GPDs.

In the factorization relations derived here, quasi gluon GPDs are related to twist-2 GPDs through 
convolutions with perturbative coefficient functions. The functions are given at one-loop accuracy. If one has 
results of quasi gluon GPDs from Lattice QCD, one can convert them to obtain gluon GPDs. So far there are no published results for quasi gluon GPDs. For quasi quark GPDs there are already some results from Lattice QCD in \cite{CLZ,LGPD,Lin1,Lin2}.
In a kinematical limit,  GPDs and quasi GPDs become corresponding PDFs and quasi PDFs, respectively. 
Taking the limit, we obtain from our results the factorization relation of quasi gluon PDFs. The factorization 
relations between quasi PDFs and twist-2 PDFs have been studied extensively in \cite{M1,M2,WZZ,STYZ,IJJSZ,WZZZ,CWZ,LMQ}.

Our work is organized as in the following: In Sect.2 we give definitions of twist-2 gluon GPDs and corresponding quasi GPDs.  The factorization relations at tree-level are derived. In Sect.3 we study the relations at one-loop.  In this section we first show how to derive gauge-invariant results, then we give 
the contributions from gluon- and quark GPDs. In Sect.4 we give our complete results for the 
factorization relations of gluon quasi GPDs. We also present the results for quasi gluon PDFs by taking 
the kinematical limit.  Sect.5 is our summary.

\par\vskip10pt

\noindent
{\bf 2.  Definitions and Factorization at Tree-Level} 

\par\vskip5pt

It is convenient to give definitions of gluon GPDs in the light-cone coordinate system, in which a vector
$a^\mu$ is expressed as $a^\mu = (a^+, a^-, \vec a_\perp) =((a^0+a^3)/\sqrt{2}, (a^0-a^3)/\sqrt{2}, a^1, a^2)$ and $a_\perp^2
=(a^1)^2+(a^2)^2$.  We introduce two light-cone vectors 
$l^\mu=(1,0,0,0)$ and $n^\mu=(0,1,0,0)$. The metric and totally antisymmetric tensor in the transverse space are given by:
\begin{equation} 
   g_\perp^{\mu\nu} = g^{\mu\nu} - l^\mu n^\nu -l^\nu n^\mu, \quad \epsilon_\perp^{\mu\nu} = \epsilon^{\sigma\rho\mu\nu} l_\sigma n_\rho, 
\end{equation} 
with $\epsilon^{0123} =1$. 
   
\par

In the coordinate system we consider an initial hadron through scattering of certain operators into 
a final state. The hadron has momentum $p$ in the initial state and $p'$ in the final state. The initial- or final hadron moves closely along light-cone direction $l$, i.e., the $z$-components of momenta are large.  We will use the following notations:
\begin{equation}
  P^\mu = \frac{1}{2}( p^\mu + p'^{\mu}) , \quad  \Delta^\mu = p'^{\mu} -p^\mu, \quad t =\Delta^2, 
  \quad \xi =  - \frac{\Delta^+} {2 P^+} =\frac{ p^{+} -p'^+} {p^{+ } + p'^+ },   
\end{equation}  
 and the gauge link: 
\begin{equation} 
   {\mathcal L}_n (y)  =  P\exp\biggr \{ -i g_s \int_0^\infty d \xi n\cdot G(\xi n + y) \biggr \}. 
\end{equation} 
The gauge link is defined in $SU(N_c)$ adjoint representation. We introduce 
\begin{equation} 
      \bar G^{+\mu}(x) = {\mathcal L}_n (x) G^{+\mu} (x), \quad 
       \bar G^{^* +\mu}(x) =  G^{+\mu} (x){\mathcal L}_n^\dagger (x), 
\label{OPG}         
\end{equation}
for our convenience. $G^{\mu\nu}$ is the field strength tensor of the gluon field $G^\mu$.       
\par 
The twist-2 gluon GPDs are defined as:  
\begin{eqnarray} 
 F_g^{\mu\nu}  (x,\xi,t)  &=&  \frac{1}{ P^+} \int\frac{d\lambda}{2\pi} e^{ i x  P^+ \lambda} 
    \langle p' \vert   \bar G^{* a, +\mu} (-\frac{1}{2} \lambda n)   \bar G^{a, +\nu}  (\frac{1}{2} \lambda n ) \vert p \rangle 
\nonumber\\
        &=& \frac{1}{2} g_\perp^{\mu\nu} F_{gU}(x,\xi,t) - \frac{i}{2} \epsilon_\perp^{\mu\nu} F_{gL}(x,\xi,t) + F_{qT}^{\mu\nu} (x,\xi,t), 
\label{GGPD}                
\end{eqnarray}    
where $\mu$ and $\nu$ are transverse indices. The tensor $F_g^{\mu\nu}$ is decomposed into 
its trace part,  antisymmetric- and trace-less symmetric part as given in the above. Each individual part can be parametrized with scalar functions for a hadron with given quantum numbers. 
For a proton or spin-1/2 hadron the parameterization is given by:
\begin{eqnarray} 
  F_{gU} (x,\xi,t) &=& \frac{1}{ P^+} \int\frac{d\lambda}{2\pi} e^{ i x P^+ \lambda} 
    \langle p' \vert g_{\perp\mu\nu}   \bar G^{* a, +\mu} (-\frac{1}{2} \lambda n)   \bar G^{a, +\nu}  (\frac{1}{2} \lambda n ) \vert p \rangle  
\nonumber\\
     &=& -\frac{1}{2  P^+} \bar u(p') \biggr [ \gamma^+ H_g (x, \xi, t) + \frac{i\sigma^{+\alpha}} {2 m} \Delta_\alpha 
     E_g (x,\xi,t) \biggr ] u(p), 
\nonumber\\
  F_{gL}(x,\xi,t)     &=& \frac{i}{ P^+} \int\frac{d\lambda}{2\pi} e^{ i x  P^+ \lambda} 
    \langle p' \vert \epsilon_{\perp\mu\nu}  \bar G^{* a, +\mu} (-\frac{1}{2} \lambda n)   \bar G^{a, +\nu}  (\frac{1}{2} \lambda n ) \vert p \rangle
\nonumber\\
     &=& -\frac{1}{2  P^+} \bar u(p') \biggr [ \gamma^+ \gamma_5  H_{gL}  (x, \xi, t) + \frac{\gamma_5 \Delta^+ } {2 m} 
       E_{gL} (x,\xi,t) \biggr ] u(p),
\nonumber\\
  F^{\mu\nu }_{gT}(x,\xi,t)  &=&  \frac{1}{2  P^+} \int\frac{d\lambda}{2\pi} e^{ i x  P^+ \lambda} 
    \langle p' \vert  {\bf S} \biggr (    \bar G^{* a, +\mu} (-\frac{1}{2} \lambda n)  \bar G^{a, +\nu}  (\frac{1}{2} \lambda n )\biggr )   \vert p \rangle  
\nonumber\\
    &=& -\frac{1}{2 P^+} {\bf S} \biggr \{ \frac{ P^+ \Delta^\nu - \Delta^+ P^\nu}{2 m  P^+} \bar u(p') \biggr [ 
           i\sigma^{+\mu} H_{gT} (x,\xi,t) +\frac{ P^+ \Delta^\mu -\Delta^+ P^\mu}{m^2} \tilde H_{gT} (x,\xi,t) 
\nonumber\\
    && +\frac{\gamma^+ \Delta^\mu -\Delta^+ \gamma^\mu}{2 m} E_{gT} (x,\xi,t) + \frac{\gamma^+ P^\mu
       -P^+ \gamma^\mu }{m} 
     \tilde E_{gT} (x,\xi,t) \biggr ] u(p)  \biggr \},  
\label{GT1}                             
\end{eqnarray}  
where the notation ${\bf S}(\cdots)$ implies that the tensors in $(\cdots)$ are symmetric and traceless.  There are in total eight twist-2 gluon GPDs. Their properties can be found in \cite{MDI, BeRa}.    

The defined gluon GPDs are nonperturbative. Since the GPDs depend on the time $t$ explicitly, they cannot be calculated directly with Lattice QCD formulated in Euclidian space-time. The new idea  is to introduce the so-called quasi GPDs\cite{Ji}. These quasi GPDs are defined with products of operators separated only 
in spatial space. Hence they 
 can be calculated with Lattice QCD directly. 
To introduce quasi GPDs, we work in cartesian coordinate system and introduce a vector
$n_z^\mu =(0,0,0,-1)$ pointing the $-z$-direction. $n_z^2 =-1$. To give the definition of quasi gluon GPDs we introduce the gauge link along the $n_z$-direction and the following notations:
\begin{eqnarray}
{\mathcal L}_z (y,\infty) &=& P\exp\biggr \{ -i g_s \int_0^\infty d \xi n_z\cdot G(\xi n_z + y) \biggr \},                    
\nonumber\\ 
      \tilde G^{z \mu}(x) &=&  {\mathcal L}_z (x) G^{z \mu} (x), \quad 
       \tilde G^{^* z \mu}(x) =  G^{z \mu} (x){\mathcal L}_z^\dagger (x).   
\end{eqnarray} 
With these notations the quasi gluon GPDs are defined as 
\begin{eqnarray} 
 {\mathcal F}_g^{\mu\nu}  (z,\xi,t)  &=&  \frac{1}{ P_z} \int\frac{d\lambda}{2\pi} e^{ i z  P_z \lambda} 
    \langle p' \vert   \tilde G^{* a, z\mu} (-\frac{1}{2} \lambda n_z ) \tilde G^{a, z \nu}  (\frac{1}{2} \lambda n_z ) \vert p \rangle 
\nonumber\\
        &=& \frac{1}{2} g_\perp^{\mu\nu} {\mathcal F}_{gU}(z,\xi,t) - \frac{i}{2} \epsilon_\perp^{\mu\nu} {\mathcal F}_{gL}(z,\xi,t) + {\mathcal F} _{qT}^{\mu\nu} (z,\xi,t),  
\label{QGGPD}               
\end{eqnarray}    
with $\mu$ and $\nu$ being transverse indices. $P_z$ is the third component of $P^\mu$, i.e., $P_z=P^3$. 
The operators in the definition are only separated in the $z$-direction. Their product does not depend 
on the time $t$. 
In the above we have decomposed the tensor ${\mathcal F}_g^{\mu\nu}$  into 
its trace part,  antisymmetric- and trace-less symmetric part similar to Eq.(\ref{GGPD}).  Each part and its parametrization for a proton or spin-1/2 hadron  is given by:
\begin{eqnarray} 
  {\mathcal F}_{gU} (z,\xi,t) &=& \frac{1}{ P_z} \int\frac{d\lambda}{2\pi} e^{ i z  P_z \lambda} 
    \langle p' \vert g_{\perp\mu\nu} \tilde G^{*z \mu} (-\frac{1}{2} \lambda n_z ) \tilde G^{z \nu}  (\frac{1}{2} \lambda n_z ) \vert p \rangle  
\nonumber\\
     &=& -\frac{1}{2  P_z} \bar u(p') \biggr [ \gamma^z {\mathcal H} _g (z, \xi, t) + \frac{i\sigma^{z\alpha}} {2 m}  \Delta_\alpha 
     {\mathcal E} _g (z,\xi,t) \biggr ] u(p), 
\nonumber\\
  {\mathcal F}_{gL}(z,\xi,t)     &=& \frac{i}{ P_z} \int\frac{d\lambda}{2\pi} e^{ i z P_z \lambda} 
    \langle p' \vert \epsilon_{\perp\mu\nu}\tilde G^{*z \mu} (-\frac{1}{2} \lambda n_z ) \tilde G^{z \nu}  (\frac{1}{2} \lambda n_z )\vert p \rangle
\nonumber\\
     &=& -\frac{1}{2  P_z} \bar u(p') \biggr [ \gamma^z \gamma_5  {\mathcal H}_{gL}  (z, \xi, t) + \frac{\gamma_5 \Delta^z } {2 m} 
       {\mathcal E} _{gL} (z,\xi,t) \biggr ] u(p),
\nonumber\\
  {\mathcal F} ^{\mu\nu }_{gT}(z,\xi,t)  &=&  \frac{1}{2  P_z} \int\frac{d\lambda}{2\pi} e^{ i z P_z \lambda} 
    \langle p' \vert  {\bf S} \biggr (   \tilde G^{*z \mu} (-\frac{1}{2} \lambda n_z ) \tilde G^{z \nu}  (\frac{1}{2} \lambda n_z ) \biggr )   \vert p \rangle  
\nonumber\\
    &=& -\frac{1}{2 P_z} {\bf S} \biggr \{ \frac{ P_z \Delta^\nu - \Delta^z P^\nu}{2 m  P_z} \bar u(p') \biggr [ 
           i\sigma^{z\mu} {\mathcal H} _{gT} (z,\xi,t) +\frac{ P_z \Delta^\mu -\Delta^z P^\mu}{m^2} \tilde {\mathcal H} _{gT} (z,\xi,t) 
\nonumber\\
    && +\frac{\gamma^z \Delta^\mu -\Delta^z \gamma^\mu}{2 m} {\mathcal E} _{gT} (z,\xi,t) + \frac{\gamma^z P^\mu
       -P_z \gamma^\mu }{m} 
     \tilde {\mathcal E} _{gT} (z,\xi,t) \biggr ] u(p)  \biggr \}.  
\label{GT2}                             
\end{eqnarray}
There are eight quasi gluon GPDs corresponding to the eight GPDs in Eq.(\ref{GT1}).   
These quasi gluon GPDs can be calculated with Lattice QCD directly because  the matrix elements have no time-dependence. 

\begin{figure}[hbt]
\begin{center}
\includegraphics[width=16cm]{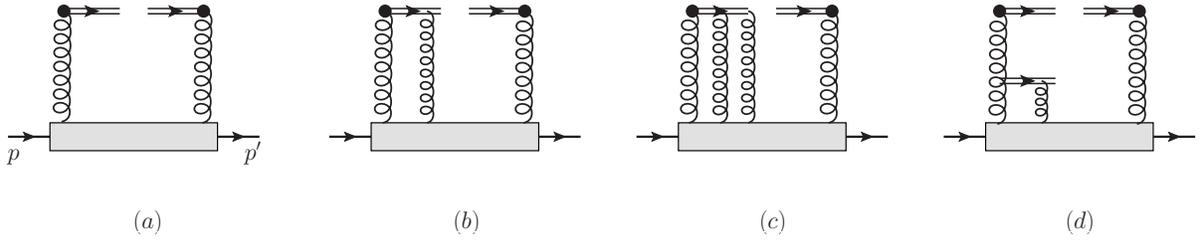}
\end{center}
\caption{The tree-level diagrams for quasi gluon GPDs. The double lines at the top of diagrams stand for gauge links along the $z$-direction.  In (d) the double line in the middle is for the gauge link along the 
light-cone $n$-direction.  } 
\label{Fig1}
\end{figure}

It is expected that the gluon GPDs and quasi gluon GPDs contain the same effects of long distance. They are related to each other. 
In the limit of large $P_z$ or $P^+$,  the relations can be calculated with perturbative theory, or  quasi GPDs can be factorized with GPDs, where perturbative coefficient functions are free from any soft divergences. 
At the leading power of the inverse of $P_z$ only twist-2 GPDs are involved in the factorization. Contributions from parton GPDs at higher twists are suppressed by the inverse of $P_z$. 
With the relations, one can obtain twist-2 parton GPDs from quasi parton GPDs calculated with Lattice QCD.

At tree-level, the factorization is derived from diagrams given in Fig.1.  Each diagram can be divided into an 
upper- and lower part. The upper part is a standard Feynman diagram in which black dots denote the insertion of the two operators in Eq.(\ref{OPG}) used to define quasi gluon GPDs.  
The lower part is the Fourier transformed hadron matrix elements of gluon fields, represented by the grey box. 
We first consider the contribution from Fig.1a and use it as an example to explain our approach. 
In Fig.1a two gluons are exchanged between the upper- and lower part. 
The contribution is: 
\begin{eqnarray} 
 {\mathcal F}^{\mu\nu}_g (z, \xi, t) ) \biggr\vert_{1a}  &=& \frac{1}{P_z} \int d^4 k \delta ( z P^3 -k^3)  
 \biggr ( i ( k_1\cdot n_z  g^{\nu\sigma} - k_1^\nu n_z^\sigma) \delta^{ac} \biggr ) 
 \nonumber\\  
 && \biggr (- i(k_2\cdot n_z  g^{\mu\rho} - k_2^\mu n_z^\rho) \delta^{ab} \biggr )  
  \int \frac{ d^4 y }{ (2\pi)^4} e^{i y\cdot k}   
  \langle p'  \vert
                G^{b}_\rho  (-\frac{y}{2}  )    
                  G^{c }_\sigma ( \frac{y}{2}  )\vert p \rangle, 
\label{F1A}                   
\end{eqnarray}                   
where the left gluon-line carries the momentum $k_1= k-\frac{1}{2} \Delta$, the right gluon-line carries the  momentum $k_2 = k+ \frac{1}{2} \Delta$.   For the total contribution from two-gluon exchanges one also needs 
to add the contribution from the diagram obtained from Fig.1a by interchanging the two gluon lines.  This causes a double counting because the Bose-symmetry is already maintained in the lower part. Hence, we need only to consider the contribution from Fig.1a. One can  add the contribution from the diagram
by interchanging the two gluon lines, 
but then one needs to divide the total contribution by a factor of 2 to avoid double-counting.

In the limit of large $P^+$ or large $P_z$  
 it is expected that $k$ scales as collinear  to the hadron momenta. We will work in Feynman gauge. In this gauge 
the gauge field $G^\mu$ in the matrix element also scales like its momentum.  Hence, we have the following 
power counting:   
\begin{equation} 
   k^\mu = (k^+,k^-, k^1, k^2) \sim (1,\lambda^2, \lambda, \lambda), \quad 
  G^\mu = (G^+,G^-, G^1, G^2) \sim (1,\lambda^2, \lambda, \lambda),    
\label{PC}    
\end{equation}
where $\lambda$ is a small parameter. 
The momentum $\Delta$ scales as the same pattern from kinematical restriction. 
An expansion in $\lambda$, called as collinear expansion, can be made. 
The leading order of the contribution from Fig.1a is at $\lambda^0$.  It is noted that the states give a power of $\lambda^{-2}$ 
because of the normalization $\langle p' \vert p\rangle = (2\pi)^3 2 p^0 \delta^3(p-p')$. It is straightforward to obtain the contribution from Fig.1a at the leading power of $\lambda$: 
\begin{eqnarray} 
 {\mathcal F}^{\mu\nu}_g (z, \xi, t) \biggr\vert_{1a}  = \frac{1}{P^+ } \int \frac{ d \lambda   }{2\pi} e^{i z \lambda   P^+ }    
   \langle p'  \vert
                G^{a,+\mu}  (-\frac{\lambda}{2} n   )      
                  G^{a,+ \nu} ( \frac{\lambda}{2} n  ) \vert p \rangle  +\cdots, 
\end{eqnarray}  
where $\cdots$ stand for power-suppressed contributions. 

Besides the contribution from Fig.1a, there are contributions from diagrams like Fig.1b, 1c, etc.,  where there are in addition to two-gluon exchange one-, two- and more exchanged gluons.  The leading contributions from diagrams with these additionally exchanged gluons are given when these gluons carry momenta only in the $+$-direction and are polarized in the $+$-direction, i.e., the corresponding gluon fields in the hadron matrix element  are all $G^+$'s. These leading contributions can be easily found. E.g.,  the leading contribution from Fig.1b,  where an extra gluon is emitted from the gauge link along the $n_z$-direction,   can be represented with Fig.1d, where this extra gluon is from the gauge link along the $n$-direction.  
The leading contributions from exchanges of one-,  two- and more gluons in addition to the two gluon lines 
in Fig,1a can  be summed into gauge links along the $n$-direction. This results in that the operator in 
the hadron matrix element is exactly that used to defined gluon GPDs in Eq.(\ref{GGPD}).  
Therefore, we have the factorization relation at tree-level:
\begin{equation} 
     {\mathcal F}^{\mu\nu}_g (z,\xi, t) =  F^{\mu\nu}_g(z, \xi, t) + \cdots, 
\label{TRE}      
\end{equation} 
 where $\cdots$ stand for contributions suppressed by positive powers of $\lambda$.

\par\vskip20pt

\noindent
{\bf 3. Factorization at One-Loop Level} 

\par\vskip5pt
In this section we study the one-loop factorization of quasi gluon GPDs. Unlike the factorization of 
quasi quark GPDs,  where it is relatively easy to maintain the gauge invariance at the leading order of the collinear expansion,  it is not straightforward to obtain gauge-invariant results in the factorization of quasi gluon GPDs beyond tree-level.  In Feynman gauge there are in the collinear expansion super-leading-power
contributions pointed out in \cite{JCTR}. There is also a problem of gluons with unphysical polarizations.  
In this section, we will first discuss the problems of super-leading-power contributions and gauge invariance
in subsection 3.1.,
then we give detailed results about gluon contributions in subsection 3.2. and those about quark contributions in subsection 3.3.

\par\vskip5pt
\noindent 
{\bf 3.1. Gauge Invariance and Super-Leading-Power Contributions} 

\par\vskip5pt 
The one-loop contributions to quasi gluon GPDs come from diagrams in Fig.2 and Fig.3. The contributions from Fig.2 may be called as real part, while the contributions from Fig.3 may be called as virtual part which is proportional to 
tree-level results.  The total contribution can be written in the form: 
\begin{equation} 
{\mathcal F}^{\sigma\rho} _g (z,\xi,t) = \frac{1}{P_z} \int d^4 k  \frac{1}{2} \Gamma^{ab, \sigma\rho \mu\nu} (k_1,k_2) 
\int \frac{d^4 y}{(2\pi)^4} e^{i y\cdot k} \langle p' \vert G^{b}_{\nu} (-\frac{y}{2} ) G^{a}_{\mu} (\frac{y}{2} ) \vert p\rangle, 
\label{S1} 
\end{equation}  
In Fig.2 and Fig.3, 
the left gluon line leaving the grey box carries the index $a,\mu$ and the momentum $k_1=k-\Delta/2$ flowing into the upper parts, 
while the right one  carries the index $b,\nu$ and the momentum $k_2=k+\Delta/2$ flowing into the grey boxes representing the hadron matrix elements. From each diagram 
one can obtain the corresponding crossed diagram where the two gluon lines are interchanged. 
$\Gamma$ is the sum of the upper parts of diagrams in Fig.2 and Fig.3  and those crossed diagrams but without the last one in Fig.2  involving the 
four gluon vertex.  $\Gamma$ is essentially the amputated Green's function defined with the $T$-ordered product of two gluon field operators and the operator used to define quasi GPD. 
We divide the total contribution by a factor of 2  to avoid the double counting mentioned in Sect.2. 
Besides $k_{1,2}$ $\Gamma$ also depends on $z$. We have suppressed the dependence. 
As an one-loop contribution, $\Gamma$ contains integrals of one-loop.

\begin{figure}[hbt]
\begin{center}
\includegraphics[width=14cm]{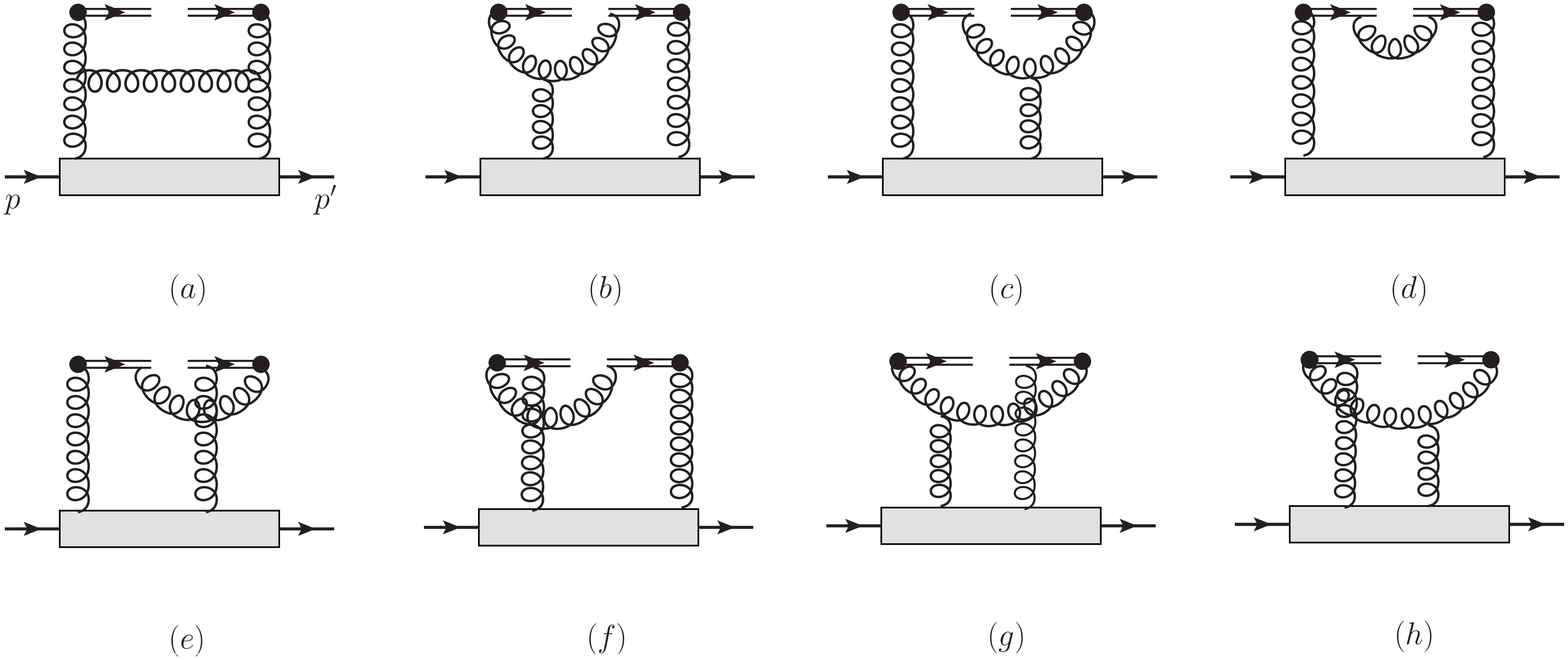}
\includegraphics[width=14cm]{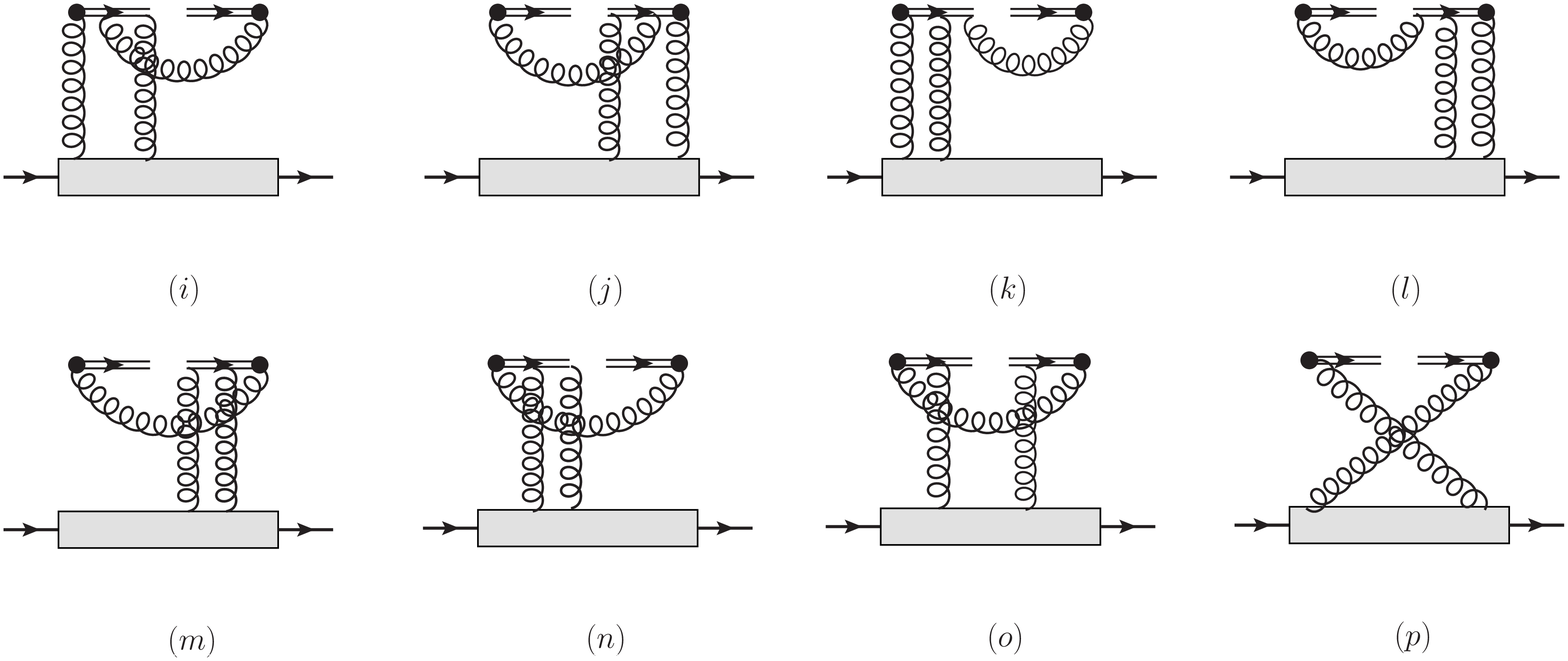}\end{center}
\caption{The one-loop diagrams for real corrections to quasi GPDs.    In the last diagram there is a four-gluon vertex in the middle.     } 
\label{Fig2a}
\end{figure}

\begin{figure}[hbt]
\begin{center}
\includegraphics[width=12cm]{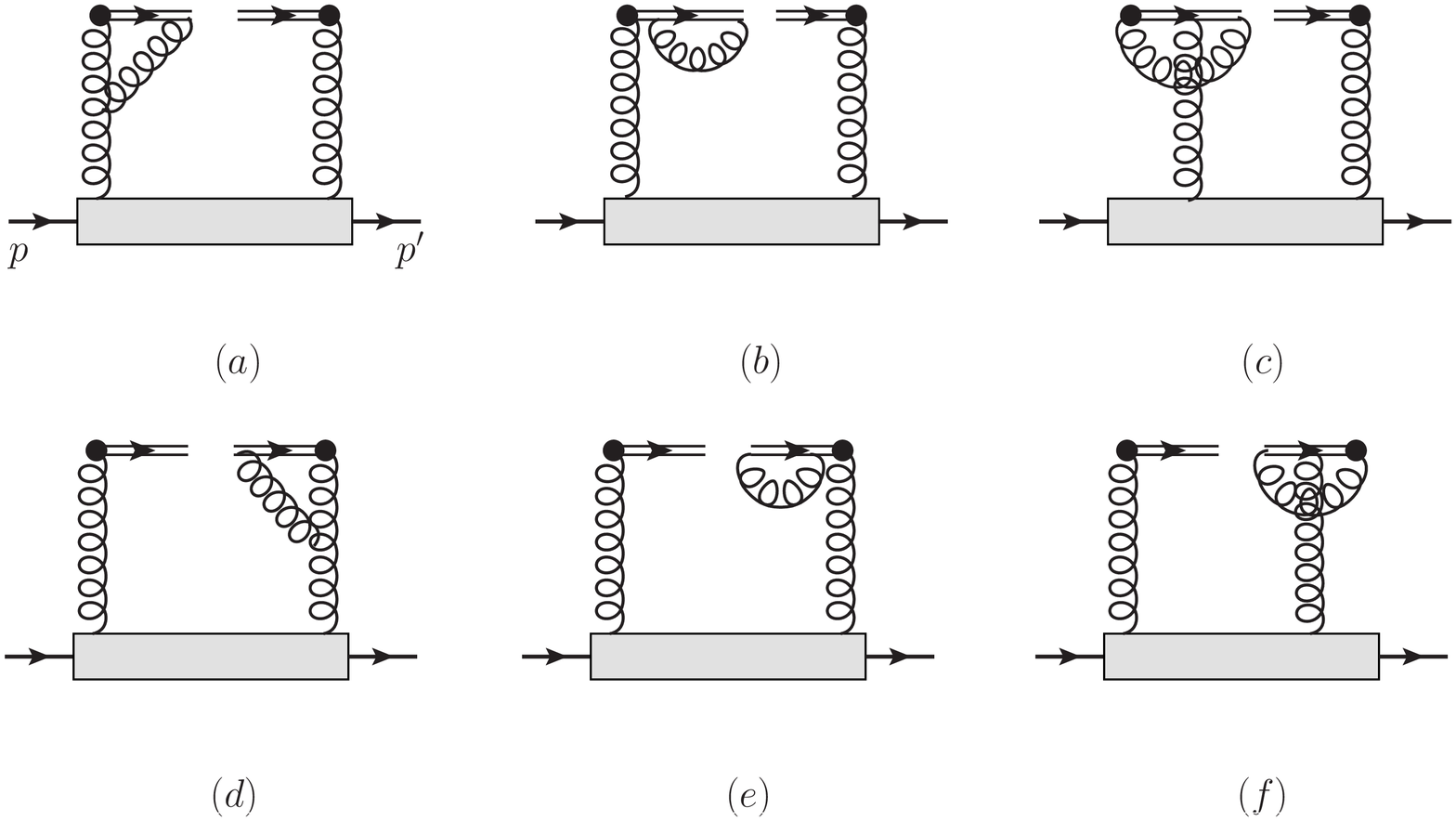}
\end{center}
\caption{The one-loop diagrams for virtual corrections to quasi GPDs.       } 
\label{Fig2b}
\end{figure}

  \par  
To avoid working with too many indices, we take the trace part of quasi gluon GPDs for our discussion here. The obtained results for the trace part ${\mathcal F}_{gU}$ also apply for other parts.    
After projecting out the trace part and using the color symmetry, the trace part is: 
\begin{equation} 
{\mathcal F}_{gU} (z,\xi,t) = \frac{1}{P_z} \int d^4 k \frac{1}{2}   {\mathcal M}_{\mu\nu} (k_1,k_2) 
\int \frac{d^4 y}{(2\pi)^4} e^{i y\cdot k} \langle p' \vert G^{c,\nu} (-\frac{y}{2} ) G^{c,\mu} (\frac{y}{2} ) \vert p\rangle, 
\label{FGU}  
\end{equation}  
with  
\begin{equation} 
 {\mathcal M}_{\mu\nu} (k_1,k_2) =  \frac{1}{N_c^2-1} g_{\perp}^{\sigma\rho} \Gamma_{\sigma\rho \mu\nu}^{ab} (k_1,k_2) \delta^{ab}. 
\end{equation} 

\par      
It is nontrivial to find the leading contributions in Feynman gauge. 
At first look one can expand ${\mathcal F}_{gU}$ given in Eq.(\ref{FGU}) in $\lambda$ straightforwardly. 
In the first step, we need to expand ${\mathcal M}^{\mu\nu}$ around $k_1=\hat k_1$ and 
$k_2=\hat k_2$: 
\begin{equation} 
 {\mathcal M}^{\mu\nu} (k_1,k_2) =  {\mathcal M}^{\mu\nu} (\hat k_1, \hat k_2) + k_{1\perp}^\sigma 
  \frac{\partial {\mathcal M}^{\mu\nu}}{\partial k_{1\perp}^\sigma} (\hat k_1,\hat k_2) 
  + k_{2\perp}^\sigma 
  \frac{\partial {\mathcal M}^{\mu\nu}}{\partial k_{2\perp}^\sigma} (\hat k_1,\hat k_2) +\cdots,
 \end{equation} 
 with $ \hat k_1^\mu =(k_1^+, 0,0,0)$ and $\hat k_2^\mu = (k_2^+, 0,0,0)$.   
With the power 
counting in Eq.(\ref{PC}) one can determine the relative importance of each term. 
It is easy to find the leading order contribution which is given only by one term:  
\begin{equation} 
  {\mathcal F}_{gU}  (z,\xi,t) = \frac{1}{2 P_z} \int d^4 k  {\mathcal M}^{--} (\hat k_1,\hat k_2)  
\int \frac{d^4 y}{(2\pi)^4} e^{i y\cdot k} \langle p' \vert G^{c,+} (-\frac{y}{2} ) G^{c,+} (\frac{y}{2} ) \vert p\rangle
  +\cdots, 
\end{equation} 
where $\cdots$ stand for contributions suppressed by power of $\lambda$. With the power counting in Eq.(\ref{PC}) and the power counting of the states, we find that this leading order contribution is at the order of $\lambda^{-2}$. The next-to-leading order is of $\lambda^{-1}$. It is expected that the leading order of physical results is $\lambda^0$. The contributions at the order of negative powers of $\lambda$ are called 
as super-leading power contributions. These contributions exist in Feynman gauge as pointed out and discussed first in \cite{JCTR}. These super-leading power contributions are gauge-variant. 
They vanish in a physical gauge. 
In order to obtain physical and gauge-invariant results in Feynman gauge, one needs to show in the first step 
that the super-leading-power contributions are in fact zero.

\par 
The Green's function, which determines ${\mathcal M}$ after amputating the external legs, is given by:
\begin{eqnarray} 
  \frac{1}{N_c^2-1} \int d^4 x_1 d^4 x_2 e^{-i k_1\cdot x_1 + ik_2\cdot x_2}  \langle 0 \vert T \biggr (  G^{a,\nu} (x_2)  G^{a,\mu} (x_1) {\mathcal O} \biggr ) \vert 0 \rangle 
\label{AMG}   
\end{eqnarray} 
with the operator 
${\mathcal O}$ used to define the quasi gluon GPDs ${\mathcal F}_{gU}$, i.e., 
\begin{equation} 
{\mathcal O}    =   \frac{1}{ P_z} \int\frac{d\lambda}{2\pi} e^{ i z  P_z \lambda} 
      \tilde G^{* a, z\mu} (-\frac{1}{2} \lambda n_z )  \tilde G^{a, z \nu}  (\frac{1}{2} \lambda n_z )g_{\perp\mu\nu}.  
\end{equation} 
For simplicity we have suppressed the dependence of all possible variables of ${\mathcal O}$. It is noted that  ${\mathcal M}$ is given by the connected contribution of the amputated Green's function. There is a disconnected contribution of the Green's function, which  involves the gluon propagator from the contraction of the two gluon fields 
explicitly given in Eq.(\ref{AMG}) and the vacuum expectation value of ${\mathcal O}$.   It is noted that the disconnected contribution is proportional to $\delta^4 (k_1-k_2)$.  Therefore, for $k_1\neq k_2$ there is no  
disconnected contribution and ${\mathcal M}$ is the amputated Green's function. 

In covariant- or Feynman gauge there is the following Ward identity:
\begin{equation} 
    k_{1\mu} k_{2\nu} {\mathcal M}^{\mu\nu} (k_1,k_2) =0. 
\label{WI}     
\end{equation} 
for $k_1\neq k_2$.  In fact $k_{1\mu} k_{2\nu} {\mathcal M}^{\mu\nu}$ is given by two terms  proportional
to $\delta^4 (k_1-k_2)$.  One is from the contracting $k_1^{\mu} k_2^\nu$ with the Green's function in Eq.(\ref{AMG}), while another is from the disconnected contribution. 
The sum of the two terms is in fact zero.  
Therefore, the identity also holds in the case of $k_1 = k_2$. 
In the following  discussion we will only take  $k_1\neq k_2$  to avoid paying attentions
to these similar $\delta^4(k_1-k_2)$-contributions for the simplicity.  However, 
the conclusion made in this subsection holds also the case of $k_1= k_2$ because ${\mathcal M}^{\mu\nu} (k_1,k_2)$ is not singular in the limit of $k_1\to k_2$, i.e., the forward limit.

We have explicitly checked the identity of ${\mathcal M}$ at one-loop level by adding all contributions  from Fig.2 and Fig.3, and found that the identity holds.     
From this identity one has by setting $k_1=\hat k_1$ and $k_2=\hat k_2$: 
\begin{equation} 
  {\mathcal M}^{--} (\hat k_1, \hat k_2) =0.  
\end{equation}       
Therefore, the super leading power contribution at order of $\lambda^{-2}$ is zero. 
The super leading power contribution at the order of $\lambda^{-1}$ has more than one term. 
From the Ward identity and the Lorentz covariance one is able to show that it vanishes too. 
Hence, the real leading-order contribution is at order of $\lambda^0$. However, with the direct expansion in $\lambda$ there are many terms at the order. It is difficult to find that the final results are gauge-invariant.  

For solving the difficulty it is useful to use Grammer-Yennie decomposition\cite{GrYe}.  We decompose the contraction of the Lorentz index $\mu$ and $\nu$ in Eq.(\ref{FGU}) as:
\begin{eqnarray} 
{\mathcal F}_{gU} (z,\xi,t) &=&  \frac{1}{2 P_z} \int d^4 k   {\mathcal M}^{\mu\nu} (k_1,k_2) 
\biggr (g_{\mu\alpha} - \frac {k_{1\mu} n_\alpha}{n\cdot k_1}  
+ \frac {k_{1\mu}  n_\alpha}{n\cdot k_1} \biggr )  \biggr (g_{\nu\beta } - \frac {k_{2\nu} n_\beta }{n\cdot k_2 } 
+ \frac {k_{2\nu} n_\beta }{n\cdot k_2} \biggr ) 
\nonumber\\
   && 
\int \frac{d^4 y}{(2\pi)^4} e^{i y\cdot k} \langle p' \vert G^{c,\beta} (-\frac{y}{2} ) G^{c,\alpha} (\frac{y}{2} ) \vert p\rangle.  
\end{eqnarray}  
 With the decomposition we write ${\mathcal F}_{gU}$ as the sum without using the Ward idenity:
 \begin{equation} 
{\mathcal F}_{gU} (z,\xi,t) = {\mathcal F}^{(-2)}_{gU} (z,\xi,t) +{\mathcal F}^{(-1)}_{gU} (z,\xi,t)+ 
{\mathcal F}^{(0)}_{gU} (z,\xi,t), 
\end{equation} 
where the first, second- and third contribution start to be nonzero at the order $\lambda^{-2}$, 
$\lambda^{-1}$ and $\lambda^{0}$, respectively. They are given by: 
\begin{eqnarray} 
{\mathcal F}_{gU}^{(-2)}  (z,\xi,t) &=&  \frac{1}{2 P_z} \int d^4 k   {\mathcal M}^{\mu\nu} (k_1,k_2) 
  \frac {k_{1\mu}  n_\alpha}{n\cdot k_1}  \frac {k_{2\nu} n_\beta }{n\cdot k_2} \int \frac{d^4 y}{(2\pi)^4} e^{i y\cdot k} \langle p' \vert G^{c,\beta} (-\frac{y}{2} ) G^{c,\alpha} (\frac{y}{2} ) \vert p\rangle, 
\nonumber\\
 {\mathcal F}_{gU}^{(-1)}  (z,\xi,t) &=&  \frac{1}{2 P_z} \int d^4 k  {\mathcal M}^{\mu\nu} (k_1,k_2) 
\biggr [  
 \frac {k_{1\mu}  n_\alpha}{n\cdot k_1}   \biggr (g_{\nu\beta } - \frac {k_{2\nu} n_\beta }{n\cdot k_2 } 
 \biggr ) + \biggr (g_{\mu\alpha} - \frac {k_{1\mu} n_\alpha}{n\cdot k_1}  
 \biggr )   \frac {k_{2\nu} n_\beta }{n\cdot k_2} \biggr ]  
\nonumber\\
   && 
\int \frac{d^4 y}{(2\pi)^4} e^{i y\cdot k} \langle p' \vert G^{c,\beta} (-\frac{y}{2} ) G^{c,\alpha} (\frac{y}{2} ) \vert p\rangle, 
\nonumber\\
{\mathcal F}_{gU}^{(0)}  (z,\xi,t) &=&  \frac{1}{2 P_z} \int d^4 k    {\mathcal M}^{\mu\nu} (k_1,k_2) 
\biggr (g_{\mu\alpha} - \frac {k_{1\mu} n_\alpha}{n\cdot k_1}  
 \biggr )  \biggr (g_{\nu\beta } - \frac {k_{2\nu} n_\beta }{n\cdot k_2 } 
 \biggr ) 
\nonumber\\
   && 
\int \frac{d^4 y}{(2\pi)^4} e^{i y\cdot k} \langle p' \vert G^{c,\beta} (-\frac{y}{2} ) G^{c,\alpha} (\frac{y}{2} ) \vert p\rangle.  
\label{GRYE}           
\end{eqnarray}  
 In the above, the first contribution given in the first line is zero because of the Ward identity in Eq.(\ref{WI}). 
 But there is no obvious reason that the second contribution in the second line is zero. 
 Before we discuss about the second term, we pointed out 
 that the last contribution ${\mathcal F}_{gU}^{(0)}$ takes a gauge invariant form because that 
 one can write it into a form with the hadron matrix element of the field strength tensor operators. 
  
Using Bose symmetry of ${\mathcal M}^{\mu\nu}$ which is 
\begin{equation} 
   {\mathcal M}^{\mu\nu} (k_1,k_2) = {\mathcal M}^{\nu\mu} ( -k_2, -k_1),  
\end{equation}        
we can write the second contribution as: 
\begin{eqnarray} 
{\mathcal F}_{gU}^{(-1)}  (z,\xi,t) = \frac{1}{ P_z} \int d^4 k  \frac { n_\alpha}{n\cdot k_1} k_{1\mu}   {\mathcal M}^{\mu\nu} (k_1,k_2)   \biggr (g_{\nu\beta } - \frac {k_{2\nu} n_\beta }{n\cdot k_2 }  \biggr )  
\int \frac{d^4 y}{(2\pi)^4} e^{i y\cdot k} \langle p' \vert G^{c,\beta}  (-\frac{y}{2} ) G^{c,\alpha} (\frac{y}{2} ) \vert p\rangle. 
\label{L2} 
\end{eqnarray} 
As pointed out earlier, ${\mathcal M}^{\mu\nu}$ is 
 the amputated Green's function in Eq.(\ref{AMG}). The quantity $k_{1\mu}   {\mathcal M}^{\mu\nu}$ is then given by:
\begin{eqnarray} 
  k_{1\mu}   {\mathcal M}^{\mu\nu} (k_1,k_2) = (-i) \frac{ ( i k_1^2 ) ( i k_2^2)}{N_c^2-1}   \int d^4 x_1 d^4 x_2 e^{-i k_1\cdot x_1 + ik_2\cdot x_2} \frac{\partial}{\partial x_1^\mu} \langle 0 \vert T \biggr (  G^{a,\nu} (x_2)  G^{a,\mu} (x_1) {\mathcal O} \biggr ) \vert 0 \rangle 
\label{K1M}   
\end{eqnarray} 
for $k_1\neq k_2$.  
The existence of the factor $k_1^2$ and $k_2^2$ is due to that we need to consider amputated Green's  function. At the order we study the amputation can be done with free field propagators.

After quantization with path integeral QCD has the symmetry under BRST transformation which is given by
\begin{eqnarray} 
  && \delta_B \psi(x) =  - i g_s C^a (x) T^a \psi(x), \quad\quad  \delta_B \bar \psi (x) = \bar \psi(x)  \left ( -  i g_s  C^a (x)  T^a  \right ),  
 \nonumber\\
 && \delta_B G^{a, \mu}(x) 
            =  (D^\mu)^{ab} C^b (x),  \quad                
 \delta_B C^a (x) = \frac{1}{2} g_s f^{abc} C^b (x) C^c(x), \quad \delta_B \bar C^a (x) = \frac{1}{\xi_G}\partial_\mu G^{a,\mu } (x),   
 \label{BRST} 
\end{eqnarray} 
 where  $C^a$ and $\bar C^a$ is the ghost- and anti-ghost field respectively. $\xi_G$ is the gauge parameter in covariant gauge. It should be taken as $1$ since we work with Feynman gauge. $D^\mu$ is the covariant derivative in the adjoint representation. 
 In order to avoid possible confusions between $T$-ordering in canonical quantization and $T^*$-ordering in path integral we will in the below use $T^*$-ordering. In our case, the difference between 
$T^*$ and $T$ can be neglected for $k_1\neq k_2$. 
We consider BRST transformation of  the Green's function:  
\begin{eqnarray} 
  0 &=&  \delta_B  \langle 0 \vert T^* \biggr ( G^{a,\nu}(x_2)  \bar C^a (x_1) {\mathcal O} \biggr ) \vert 0 \rangle
\nonumber\\
&=& \frac{1}{\xi_G} \langle 0 \vert T^* \biggr ( G^{a,\nu}(x_2)  \partial_\mu G^{a,\mu} (x_1)  {\mathcal O} \biggr ) \vert 0 \rangle 
   + \langle 0 \vert T^* \biggr (  (D^\nu)^{ab} C^b (x_2)   \bar C^a (x_1) {\mathcal O} \biggr ) \vert 0 \rangle,    
   \end{eqnarray} 
where we have used $\delta_B {\mathcal O} =0$ because ${\mathcal O}$ is gauge invariant. Therefore, with the symmetry of BRST transformation  $ k_{1\mu}   {\mathcal M}^{\mu\nu} (k_1,k_2)$ in Eq.(\ref{K1M}) is related to the following matrix element of ghost fields:  
\begin{eqnarray} 
  k_{1\mu}   {\mathcal M}^{\mu\nu} (k_1,k_2)  = i  \frac{ ( i k_1^2 ) ( i k_2^2)}{N_c^2-1}   \xi_G \int d^4 x_1 d^4 x_2 e^{-i k_1\cdot x_1 + ik_2\cdot x_2} \langle 0 \vert T^*  \biggr (  (D^\nu)^{ab}(x_2) C^b (x_2)   \bar C^a (x_1) {\mathcal O} \biggr ) \vert 0 \rangle,  
 \label{BRG}    
\end{eqnarray}  
In the above, the derivative $\partial^\nu$ in the covariant derivative  $(D^\nu)^{ab}=\partial^\mu \delta^{ab}  + g_s f^{cab} G^{c,\mu} (x) $ can be worked out through partial integrations. It gives a contribution 
proportional to $k_2^\nu$.  
Hence, with Eq.(\ref{BRG}) we write  
$k_{1\mu} {\mathcal M}^{\mu\nu}(k_1,k_2)$ into the form:
\begin{equation} 
k_{1\mu}   {\mathcal M}^{\mu\nu} (k_1,k_2) =   k_2 ^\nu  {\mathcal C}_0 (k_1,k_2) 
                  - k_2^2 {\mathcal C} ^\nu (k_1,k_2),   
\label{K11} 
\end{equation} 
where ${\mathcal C}_0$ and ${\mathcal C}^\nu$ are given by: 
\begin{eqnarray} 
    {\mathcal C}_0(k_1,k_2) &=&  \frac{ ( i k_1^2 ) ( i k_2^2)}{N_c^2-1}   \xi_G \int d^4 x_1 d^4 x_2 e^{-i k_1\cdot x_1 + ik_2\cdot x_2} \langle 0 \vert T^*  \biggr (  C^a (x_2)   \bar C^a (x_1) {\mathcal O} \biggr ) \vert 0 \rangle, 
\nonumber\\
      {\mathcal C}^\nu (k_1,k_2) &=&  \frac{ ( i k_1^2 ) \xi_G}{N_c^2-1}  \int d^4 x_1 d^4 x_2 e^{-i k_1\cdot x_1 + ik_2\cdot x_2} g_s f^{cab} \langle 0 \vert T^*  \biggr (  G^{c,\nu} (x_2)  C^b (x_2)   \bar C^a (x_1) {\mathcal O} \biggr ) \vert 0 \rangle.  
 \label{GC}    
\end{eqnarray}  
Because of the Ward identity in Eq.(\ref{WI}),  one has the relation:
\begin{equation}
  {\mathcal C}_0(k_1,k_2) =k_2\cdot {\mathcal C} (k_1,k_2). 
 \end{equation}  
This implies that the two Green's functions in Eq.(\ref{GC}) are related to each other because of gauge symmetry. 
This is also verified by explicit calculation of  the two Green's functions of ghost fields in Eq.(\ref{BRG}) at the considered order. 

Using the result in Eq.(\ref{K11}) we can write the second contribution in Eq.(\ref{GRYE}) as:
 \begin{eqnarray} 
{\mathcal F}_{gU}^{(-1)}   (z,\xi,t)  &=& \frac{-i }{P_z} \int d^4 k  \frac {1}{n\cdot k_1}  {\mathcal C}_0(k_1,k_2)    
 \int \frac{d^4 y}{(2\pi)^4} e^{i y\cdot k} \langle p' \vert   \partial_\beta G^{c,\beta}  (-\frac{y}{2} ) n\cdot G^{c} (\frac{y}{2} )
  \vert p\rangle, 
\nonumber\\
   && +  \frac{1}{ P_z} \int d^4 k  \frac {1}{n\cdot k_1} {\mathcal C}_\nu (k_1,k_2)    
 \int \frac{d^4 y}{(2\pi)^4} e^{i y\cdot k} \langle p' \vert   \partial^2 G^{c,\nu}  (-\frac{y}{2} ) n\cdot G^{c} (\frac{y}{2} ) \vert p\rangle,    
\label{L22} 
\end{eqnarray}        
where some partial integrations have been done to convert the factor $k_2^\mu$ and $k_2^2$ in Eq.(\ref{K11}) as derivatives acting on 
the corresponding fields, respectively. In the second line of Eq.(\ref{L22})  a part of EOM operator is involved. EOM  in Feynman gauge with $\xi_G=1$ reads: 
\begin{equation} 
    \partial^2 G^{a,\mu} (x) + {\mathcal O }(g_s)=0,  
\end{equation}           
where the terms at the order of $g_s$ consist of a quark-, ghost- and gluonic part. At the order we consider, we can neglect these parts. Therefore, the       
contribution in the second line of Eq.(\ref{L22}) is zero at the order we work. It is noted that in order to find the complete 
contribution of EOM operator, one needs to consider contributions beyond two-parton exchanges. In this section we only consider two-parton exchanges at the order of $g_s^2$ as those in Fig.2 and Fig.3.  
The neglected parts of EOM operators consists of quark-, ghost-  and gluon color current operators. 
We have made analyses of diagrams which have $q\bar q g$- and $c\bar c g$ three-parton exchanges. 
Indeed, we find the contributions at order of $g_s^3$ involving quark- and ghost color-current operators. Adding these contributions to that  in the last line of Eq.(\ref{L22}), the sum is related to the matrix element 
with EOM operator if one neglects the part of gluon color-current operator. To include the gluon part 
one needs to consider three-gluon- and four-gluon exchanges.

\begin{figure}[hbt]
\begin{center}
\includegraphics[width=10cm]{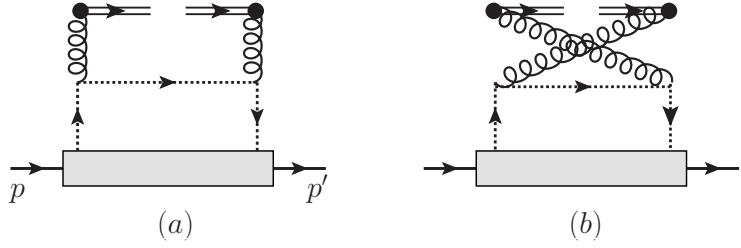}
\end{center}
\caption{The Feynman diagrams for the contributions with ghost fields. 
     } 
\label{Fig4}
\end{figure}

Unlike the one-loop factorization of quasi quark GPDs studied in \cite{MaPaZh}, where there is no contribution from ghosts, here the ghost contribution appears at the order. The contribution is given by diagrams in Fig.4. 
It is noted that the upper part of the diagrams is the amputated Green's function defined in the first line of 
Eq.(\ref{GC}).  With this fact, the contribution from Fig.4 in Feynman gauge is:
\begin{eqnarray} 
{\mathcal F}_{gU} (z,\xi,t) \biggr \vert_{Ghost} &=&  \frac{1}{P_z  } \int d^4 k  {\mathcal C}_0(k_1,k_2)  
\int \frac{d^4 y}{(2\pi)^4} e^{i y\cdot k} \langle p' \vert  \bar C^a ( -\frac{y}{2} )  C^a (\frac{y}{2})\vert p\rangle. 
\end{eqnarray} 
We can re-write the contribution as: 
 \begin{eqnarray} 
{\mathcal F}_{gU} (z,\xi,t) \biggr \vert_{Ghost}        &=&\frac{1}{ P_z} \int d^4 k  \frac{i }{n\cdot k_1}  {\mathcal C}_0(k_1,k_2)  
\int \frac{d^4 y}{(2\pi)^4} e^{i y\cdot k} \langle p' \vert  \bar C^a ( -\frac{y}{2} )  \partial^+ C^a (\frac{y}{2})\vert p\rangle
\nonumber\\
 &=&\frac{1}{P_z} \int d^4 k  \frac{i }{n\cdot k_1}  {\mathcal C}_0 (k_1,k_2)  
\int \frac{d^4 y}{(2\pi)^4} e^{i y\cdot k} \langle p' \vert  \bar C^a ( -\frac{y}{2} )  (D^+)^{ab} C^b (\frac{y}{2})\vert p\rangle, 
\end{eqnarray}  
 where in the last line we have replaced the derivative $\partial^+$ with the covariant derivative $D^+$. 
 The difference is an effect of ${\mathcal O}(g_s)$ which can be neglected at the considered order. 
Adding the ghost contribution to the nonzero contribution in Eq.(\ref{L22}), we find that 
the sum is related to the hadron matrix element of an operator which is a BRST-variation operator, i.e., 
 \begin{eqnarray} 
&& {\mathcal F}^{(-1)}_{gU} (z,\xi,t)  +  {\mathcal F}_{gU}(z,\xi,t) \biggr \vert_{Ghost}  
\nonumber\\
&& = 
 \frac{1}{ P_z} \int d^4 k  \frac{i }{n\cdot k_1}  {\mathcal C}_0(k_1,k_2)  
\int \frac{d^4 y}{(2\pi)^4} e^{i y\cdot k} \langle p' \vert   \biggr [ \bar C^a ( -\frac{y}{2} )  (n\cdot D)^{ab} C^b (\frac{y}{2})  -\partial_\beta G^{c,\beta}  (-\frac{y}{2} ) n\cdot G^{c} (\frac{y}{2} ) \biggr ]  \vert p\rangle 
\nonumber\\
 &&= - 
 \frac{1}{ P_z} \int d^4 k  \frac{i }{n\cdot k_1}  {\mathcal C}_0(k_1,k_2)  
\int \frac{d^4 y}{(2\pi)^4} e^{i y\cdot k} \langle p' \vert  \delta_B \biggr [ \bar C^a ( -\frac{y}{2} )  n\cdot G^{a} (\frac{y}{2} )  \biggr ]  \vert p\rangle,
\label{SGG}  
\end{eqnarray} 
where $\delta_B$ denotes BRST transformation given in Eq.(\ref{BRST}) with $\xi_G=1$. Because of BRST symmetry, the matrix element in the last line and hence the sum is zero. 
\par

With the results represented in the above, we conclude that after using EOM and adding the ghost contribution from Fig.4,  the one-loop gluon contribution is gauge invariant.  At the leading power or at twist-2  the quasi gluon GPDs is given by:
\begin{eqnarray} 
{\mathcal F}_{gU} (z,\xi,t) &=&  \frac{1}{\sqrt{2}  P^+} \int   \frac{ d k^+ }{k_1^+ k_2^+} {\mathcal M}_{\mu\nu} (\hat k_1, \hat k_2)  \int \frac{d y^- }{2\pi} e^{i y^- k^+ }   \langle p' \vert G^{c,+\nu}  (-\frac{y^-}{2} n ) G^{c,+\mu } (\frac{y^-}{2} n ) \vert p\rangle +\cdots,   
\end{eqnarray}   
where the index $\mu$ and $\nu$ are transverse and $\cdots$ are the power suppressed contributions. 
This result can be generalized to the contributions to ${\mathcal F}_{gL}$ and ${\mathcal F}_{gT}^{\mu\nu}$. 
Since  $\mu$ and $\nu$ are transverse, not all diagrams in Fig.2 and Fig.3 will contribute to quasi gluon GPDs.

\par
Before we go to detailed results a brief discussion may be useful.  The result in Eq.(\ref{SGG}) corresponds  
to the statement that in cross-sections of scattering with gluons the contributions from unphysically polarized gluons in cross-sections are cancelled by contributions from ghosts. A gauge-invariant operator ${\mathcal O}$ in general will be mixed with other operators because of quantum fluctuations. If ${\mathcal O}$ is local, it has been proven that the mixing pattern is given schematically as:
\begin{equation} 
   {\mathcal O} \sim [ {\mathcal O}_g ]  +  [ \delta_B {\mathcal O}_B ]  + [ {\mathcal O}_{EOM} ], 
\end{equation} 
where  $[ {\mathcal O}_g ]$,   $ [\delta_B {\mathcal O}_B ]$ and  $ [ {\mathcal O}_{EOM} ]$ denote a set of gauge-invariant operators, BRST-variation operators and those containing the EOM operator. It is understood 
that each operator in the sets is multiplied with a corresponding perturbative coefficient. When sandwiching between physical states, the last two sets of operators give no contribution. This mixing pattern  is proven for local 
operators in \cite{JoLee, Jo}. However, there is no proof for the case when the operator ${\mathcal O}$ is a nonlocal one.  Our results here show that for the nonlocal operator used to defined quasi gluon GPDs 
it is also the case.  One may think that the moments of quasi gluon GPDs are related to local operators 
and hence the nonlocal operator can be represented by local operators like gluon GPDs. However, 
it is not the case.  Since the momentum fraction $z$ of quasi GPDs is from $-\infty$ to $\infty$, the moments of quasi GPDs 
can not be defined.

\par\vskip10pt

\noindent
{\bf 3.2. One-Loop Contribution from gluon GPDs } 

\par\vskip5pt
In this section we give the gauge invariant contribution from Fig.2 and Fig.3. With the results of the last subsection, the contribution is:
\begin{eqnarray} 
 {\mathcal F}_g^{\sigma\rho}  (z,\xi,t) &=&  \frac{1}{ 2 P_z } \int   \frac{ d k^+ }{k_1^+ k_2^+}\hat \Gamma^{\sigma\rho}_{\ \ \mu\nu} (\hat k_1, \hat k_2) 
  \int \frac{d \lambda }{2\pi} e^{i \lambda k^+ }   \langle p' \vert G^{c,+\nu}  (-\frac{\lambda}{2} n ) G^{c,+\mu } (\frac{\lambda}{2} n ) \vert p\rangle,
\nonumber\\
       \hat \Gamma^{\sigma\rho\mu\nu} (\hat k_1, \hat k_2) &=& \frac{1}{N_c^2-1}  \delta^{ab} \Gamma^{ab, \sigma\rho\mu\nu} (\hat k_1, \hat k_2)  ,  
\label{ONEL}          
\end{eqnarray}   
where all indices $\mu,\nu,\sigma$ and $\rho$ are transverse and $\hat k_{1,2}$ are given by:
\begin{equation} 
   \hat k_1^\mu = ((x+\xi) P^+, 0,0,0), \quad \hat k_2^\mu = ((x-\xi)P^+,0,0,0).
\end{equation}    
For the transverse index $\mu$ and $\nu$, we only need to calculate the following diagrams: The first four- and the last diagrams in Fig.2 and the first two diagrams in the first- and second row  in Fig.3.

\par 
The calculation is straightforward in Feynman gauge.  Collinear- and U.V. divergences are regularized 
with dimensional regularization. U.V. divergences are subtracted with ${\rm \overline{MS}}$-scheme. 
The contributions from Fig.3 are proportional to 
$\delta(x-z)$. The needed contributions from Fig.3 after the U.V. subtraction are:
\begin{eqnarray} 
\hat \Gamma^{\sigma\rho\mu\nu}(\hat k_1, \hat k_2)\biggr\vert _{3a}  &=&\frac{\alpha_s}{4\pi} N_c P_z g_\perp^{\sigma\mu} g_\perp^ {\rho\nu}\delta(x-z) (z^2-\xi^2)  \biggr \{ 
        \ln (z+\xi)^2  -3   - 2 \int_{-1}^{1} d y \frac{\ln\vert y-z\vert }{\vert y-z\vert}
\nonumber\\
    && +\biggr ( \frac{2}{\epsilon_c} + \ln \frac{4\pi \mu_c^2}{(2P_z)^2 e^\gamma} \biggr )
       \biggr ( \ln (z+\xi)^2 - \ln (1-z^2) + \int_{-1}^{1} d y \frac{1}{\vert y-z\vert} \biggr ) 
\nonumber\\
    &&  - 2\ln^2 \vert z+\xi\vert + \ln^2 (1-z) + \ln^2 (1+z) +\frac{\pi^2}{3} -  \biggr ( \frac{2}{\epsilon_c} + \ln \frac{4\pi  \mu_c^2}{(2P_z)^2 e^\gamma} \biggr ) \biggr \},  
\nonumber\\
\hat \Gamma^{\sigma\rho\mu\nu}(\hat k_1, \hat k_2)\biggr\vert _{3b }  &=& \frac{\alpha_s}{4\pi} N_c P_z
g_\perp^{\sigma\mu} g_\perp^ {\rho\nu} \delta (x-z) (z^2-\xi^2) \biggr ( 2+ \ln\frac{\mu^2}{(2P_z)^2 (1-x^2)} 
   + \int_{-1}^{1} d y \frac{1}{\vert y-z\vert} \biggr ), 
\nonumber\\   
\hat \Gamma^{\sigma\rho\mu\nu}(\hat k_1, \hat k_2)\biggr\vert _{3e} &=& \hat \Gamma^{\rho\sigma\nu\mu}(\hat k_2, \hat k_1)\biggr\vert _{3b} , 
\quad 
\hat \Gamma^{\sigma\rho\mu\nu}(\hat k_1, \hat k_2)\biggr\vert _{3d} = \hat \Gamma^{\rho\sigma\nu\mu}( \hat k_2, \hat k_1)\biggr\vert _{3a},   
\end{eqnarray}  
where $1/\epsilon_c$ is the collinear pole at $\epsilon_c=4-d$ in $d$-dimensional space-time. In these results 
there are integrals of $y$ which are divergent because of the end-point singularity at $y=z$. These divergences will be cancelled by those appearing in contributions from diagrams in Fig.2.   

The contributions from Fig.2b, Fig.2c and Fig.2d are:
\begin{eqnarray} 
\hat \Gamma^{\sigma\rho\mu\nu}(\hat k_1, \hat k_2)\biggr\vert _{2b}  &=& -\frac{\alpha_s}{4\pi} N_c P_z g_\perp^{\sigma\mu} g_\perp^ {\rho\nu} (z^2-\xi^2) \biggr \{ \frac{1}{x+\xi} \biggr ( \frac{\vert z+\xi\vert}{x-z}  + \frac{\vert x-z\vert}{z+\xi}  \biggr ) +\frac{1}{2} \biggr ( \frac{2}{\epsilon_c} + \ln \frac{4\pi \mu_c^2}{(2P_z)^2 e^\gamma} \biggr )  
\nonumber\\
  && \biggr [ \frac{(x+z+2\xi)}{(x+\xi)(x-z)} \epsilon( z+\xi)  + \biggr ( \frac{2}{\vert x-z\vert} - \frac{\epsilon(x-z)}{x+\xi} \biggr ) \biggr ] 
\nonumber\\
  && - \biggr [ \frac{(x+z+2\xi)}{(x+\xi)(x-z) } \epsilon( z+\xi) \ln \vert z+\xi\vert + \biggr ( \frac{2}{\vert x-z\vert} - \frac{\epsilon(x-z)}{x+\xi} \biggr )\ln \vert x-z\vert \biggr ]  \biggr \} ,  
\nonumber\\
\hat \Gamma^{\sigma\rho\mu\nu}(\hat k_1, \hat k_2)\biggr\vert _{2c} &=& \hat \Gamma^{\rho\sigma\nu\mu}(\hat k_2, \hat k_1)\biggr\vert _{2b} , 
\nonumber\\
\hat \Gamma^{\sigma\rho\mu\nu}(\hat k_1, \hat k_2)\biggr\vert _{2d }  &=& - \frac{\alpha_s}{2\pi} N_c P_z
g_\perp^{\sigma\mu} g_\perp^ {\rho\nu} (z^2-\xi^2)  \frac{1}{\vert x-z\vert}, 
\end{eqnarray} 
where $\epsilon(z)$ is the sign function defined as:
\begin{equation} 
   \epsilon(z) = \theta(z) -\theta(-z). 
\end{equation}    
These three contributions also have the end-point singularity at $x=z$. However, when summed with the corresponding contributions from Fig.3, the singularity is cancelled we have:
\begin{eqnarray} 
\hat \Gamma^{\sigma\rho\mu\nu}(\hat k_1, \hat k_2)\biggr\vert _{2d+3b+3e} &=& 
  \frac{\alpha_s}{2\pi} N_c P_z
g_\perp^{\sigma\mu} g_\perp^ {\rho\nu} (z^2-\xi^2) \biggr [ -  \biggr ( \frac{1}{\vert x-z\vert}\biggr )_+ 
    + \delta (x-z)\biggr ( 2+ \ln\frac{\mu^2}{(2P_z)^2 (1-x^2)} \biggr)   \biggr ], 
\nonumber\\
\hat \Gamma^{\sigma\rho\mu\nu}(\hat k_1, \hat k_2)\biggr\vert _{2b +3a }  &=& -  \frac{\alpha_s}{4\pi} N_c P_z g_\perp^{\sigma\mu} g_\perp^ {\rho\nu} (z^2-\xi^2) \biggr \{ \frac{1}{x+\xi} \biggr ( \frac{\vert z+\xi\vert}{x-z}  + \frac{\vert x-z\vert}{z+\xi}  \biggr ) +\frac{1}{2} \biggr ( \frac{2}{\epsilon_c} + \ln \frac{4\pi \mu_c^2}{(2P_z)^2 e^\gamma} \biggr )  
\nonumber\\
  && \biggr [ \frac{(x+z+2\xi)}{(x+\xi)(x-z)} \epsilon( z+\xi)  + \biggr ( \frac{2}{\vert x-z\vert_+ } - \frac{\epsilon(x-z)}{x+\xi} \biggr ) \biggr ] -2 \biggr (  \frac{\ln \vert x-z\vert}{\vert x-z\vert} \biggr )_+
 \nonumber\\
  && -  \frac{(x+z+2\xi)}{(x+\xi)(x-z) } \epsilon( z+\xi) \ln \vert z+\xi\vert  +  \frac{\epsilon(x-z)\ln \vert x-z\vert}{x+\xi}  
\nonumber\\
 &&  -\delta (x-z) 
  \biggr [    \biggr ( \frac{2}{\epsilon_c} + \ln \frac{4\pi \mu_c^2}{(2P_z)^2 e^\gamma} \biggr )
        \ln \frac{(z+\xi)^2}{(1-z^2)} + \ln^2 (1-z) + \ln^2 (1+z)  
\nonumber\\
    &&  - 2\ln^2 \vert z+\xi\vert  +\frac{\pi^2}{3}-3 -  \biggr ( \frac{2}{\epsilon_c}  +\ln \frac{ 4\pi \mu_c^2}{(2P_z)^2 (z+\xi)^2 e^\gamma} \biggr ) \biggr ] \biggr \},  
\end{eqnarray}    
with the $+$-distribution defined as:
\begin{equation} 
\int_{-1}^{1} d x  \biggr ( \frac{f(x)} {\vert x-z\vert}\biggr )_+   t(x) =\int_{-1}^{1} d x \biggr [  \frac{f(x)} {\vert x-z\vert}  t(x) - \delta (x-z) t(x) \biggr ( \int_{-1}^{1} d y \frac{f(y) } {\vert y-z\vert} \biggr ) \biggr]  . 
\end{equation} 
The same cancellation also  happens in the sum of Fig.2c and Fig.3d which can be obtained from permutations from the sum of Fig.3b and Fig.3a. The contributions from Fig.2p and Fig.2a are:
\begin{eqnarray} 
\hat \Gamma^{\sigma\rho\mu\nu}(\hat k_1, \hat k_2)\biggr\vert _{2p }  &=& -\frac{\alpha_s}{4\pi} N_c P_z\frac{1}{2\xi} \biggr [\frac{1}{2} ( 2 g_\perp^{\mu\nu} g_\perp^{\sigma\rho} - g_\perp^{\mu\rho} g_\perp^{\nu\sigma}  
    -  g_\perp^{\mu\sigma} g_\perp^{\nu\rho} )(\xi^2-z^2)  \  \biggr ( \frac{2}{\epsilon_c} 
\nonumber\\     
   &&  + \ln \frac{4\pi \mu_c^2}{(2P_z(z+\xi))^2 e^\gamma} \biggr )  \epsilon(z+\xi) 
   -  g_\perp^{\mu\nu} g_\perp^{\sigma\rho}(z+\xi)^2\epsilon(z+\xi) -(\xi\to -\xi)  
   \biggr ], 
\nonumber\\
\hat \Gamma^{\sigma\rho\mu\nu}(\hat k_1, \hat k_2)\biggr\vert _{2a}  &=& \frac{\alpha_s}{4\pi} N_c P_z 
\biggr \{ -(g_\perp^{\mu\nu} g_\perp^{\sigma\rho} + g_\perp^{\mu\sigma} g_\perp^{\nu\rho} +g_\perp^{\mu\rho} g_\perp^{\nu\sigma} )\frac{1}{3(x^2-\xi^2)} \biggr [ \frac{x-\xi}{2\xi} \vert z+\xi\vert^3 + \frac{1}{2} \vert x-z\vert^3 \biggr ] 
\nonumber\\
   &&-\frac{1}{2}  \biggr ( (x^2-2\xi^2+z^2) g_\perp^{\mu\nu} g_\perp^{\sigma\rho} 
       + 2 (\xi^2 +xz) g_\perp^{\mu\sigma}g_\perp^{\nu\rho} + 2 (\xi^2-xz)  g_\perp^{\mu\rho}g_\perp^{\nu\sigma } \biggr ) 
       \nonumber\\    
   &&  \biggr [  \frac{ \vert z+\xi \vert }{\xi( x+\xi)}  
        \biggr ( \frac{2}{\epsilon_c} + \ln \frac{4\pi \mu_c^2 e^2 }{e^\gamma (2(z+\xi) P_z)^2}  \biggr ) 
 +\frac{ \vert x-z \vert }{x^2-\xi^2}  \biggr ( \frac{2}{\epsilon_c} + \ln \frac{4\pi \mu_c^2 e^2}{e^\gamma (2 (x-z) P_z)^2} \biggr )   \biggr ] 
\nonumber\\
    && +  (z^2 - \xi^2) g_\perp^{\mu\sigma}g_\perp^{\nu\rho} \biggr [  \frac{x-\xi}{4\xi (x+\xi)} \epsilon(z+\xi) \biggr ( \frac{2}{\epsilon_c} + \ln \frac{4\pi \mu_c^2}{e^\gamma (2 (z+\xi) P_z)^2}\biggr )  
\nonumber\\
    && -\frac{\epsilon(x-z)}{2(x+\xi)} \biggr ( \frac{2}{\epsilon_c} + \ln \frac{4\pi \mu_c^2}{e^\gamma (2 (x-z) P_z)^2}\biggr )     
     \biggr ]     +(\xi \to -\xi ) \biggr \}.         
\end{eqnarray}
These two contributions are free from the end-point singularity.

For the total contributions to $\hat \Gamma^{\sigma\rho\mu\nu}$ we have to add the contributions from the corresponding crossed diagrams as explained after Eq.(\ref{S1}). The total is then: 
\begin{eqnarray} 
 \hat \Gamma^{\sigma\rho\mu\nu}(\hat k_1, \hat k_2) &=&  \hat \Gamma^{\sigma\rho\mu\nu}(\hat k_1, \hat k_2)\biggr\vert _{2a+2b+2c+2d+3a+3b+3d+3e} + \hat \Gamma^{\sigma\rho\mu\nu}(\hat k_1, \hat k_2)\biggr\vert _{2p} 
 \nonumber\\
 && + \hat \Gamma^{\sigma\rho\nu\mu}(-\hat k_2, -\hat k_1 )\biggr\vert _{2a+2b+2c+2d+3a+3b+3d+3e}. 
 \label{SUMG} 
\end{eqnarray} 
There are still collinear divergences in the contribution. These divergences are due to collinear gluons in Fig.2 and Fig.3. However, they are double-counted.  In the tree-level results of Eq.(\ref{TRE}) the right-hand side  is the gluon GPDs. If we calculate the gluon GPDs at one-loop as done here for the quasi one,  the result also contains the contributions from collinear gluons. 
If we identify the Fourier-transformed hadron matrix element  as gluon GPDs in the right hand side of Eq.(\ref{ONEL})  for 
our one-loop result, then the contributions from collinear gluons in the result are already included in the tree-level result
in Eq.(\ref{TRE}). Therefore, there is a double counting for these contributions. A subtraction is needed to avoid the double-counting and to obtain correct results. 

For the subtraction we need to calculate the one-loop contribution of gluon GPDs. The contribution is represented by the same diagrams in Fig.2 and Fig.3 in which the double lines now  represent the gauge links along the light-cone $n$-direction instead of the $n_z$-direction.  We denote the contribution of the subtraction as:  
\begin{eqnarray} 
 \bar F_g^{\sigma\rho}  (z,\xi,t) &=&  \frac{1}{2  P^+} \int   \frac{ d k^+ }{k_1^+ k_2^+}\bar  \Gamma^{\sigma\rho}_{\ \ \mu\nu} (\hat k_1, \hat k_2)  
  \int \frac{d \lambda }{2\pi} e^{i \lambda k^+ }   \langle p' \vert G^{c,+\nu}  (-\frac{\lambda}{2} n ) G^{c,+\mu } (\frac{\lambda}{2} n ) \vert p\rangle.   
\end{eqnarray}   
The subtraction is done by the replacement:
\begin{equation} 
 {\mathcal F}^{\sigma\rho}_g (x,\xi,t) \to  {\mathcal F}^{\sigma\rho}_g (x,\xi,t) -  \bar F^{\sigma\rho}_g (x,\xi,t). 
\end{equation}
Since the gauge links are along the $n$-direction, the contributions from Fig.2d in Fig.2, Fig.3b and 3e in Fig.3 are zero because of $n^2=0$. Calculating the nonzero contributions  from Fig.2a, 2b, 2c and 2p in Fig.2 and Fig.3a in Fig.3, we find that these contributions contain exactly the same collinearly divergent contributions in ${\mathcal F}^{\sigma\rho}_g$, and 
the subtraction can be  effectively made by the replacement in ${\mathcal F}^{\sigma\rho}$: 
\begin{equation} 
    \frac{2}{\epsilon_c} + \ln \frac{4\pi \mu_c^2}{e^\gamma (2P_z)^2} \to \ln \frac{\mu^2}{ (2P_z)^2}. 
\label{SUBC}    
\end{equation}      
This implies that the quasi gluon GPDs contain the same 
long-distance effects in the gluon GPDs, or a factorization of quasi gluon GPDS with gluon GPDs holds. 
Performing the sum in Eq.(\ref{SUMG}) after the subtraction,  the complete contribution from gluon GPDs  
is obtained and summarized in Sect.4.

Before we turn to the one-loop contribution from quark GPDs, we briefly discuss the renormalization of the quasi gluon GPDs. 
The quasi gluon GPDs are renormalized as:
\begin{equation}
 \biggr ( {\mathcal F}_g^{\mu\nu} (z, \xi, t) \biggr )_0 = Z_3 Z_G {\mathcal F}_g^{\mu\nu} (z, \xi, t)
 +\cdots, 
\end{equation} 
where the quasi GPDs in the left side is unrenormalized one and $\cdots$ stand for possible mixings. 
$Z_3$ is the renormalization constant of gluon wave function in Feynman gauge. It is given by:
\begin{equation} 
   Z_3 = 1-\frac{\alpha_s}{4\pi} \left ( \frac{2}{\epsilon} -\gamma + \ln 4\pi \right) \biggr [ 
       -\frac{5}{3} N_c +\frac{2}{3} n_f \biggr ] + {\mathcal O}(\alpha_s^2),  
\end{equation} 
with $n_f$ as the number of quark flavors. $Z_G$ is determined only by the contributions from Fig.3b and 3e, because that only these two diagrams have the U.V. divergences with the pole at $d=4$. We obtain: 
\begin{equation} 
   Z_G = 1+\frac{\alpha_s N_c}{2\pi} \left ( \frac{2}{\epsilon} -\gamma + \ln 4\pi \right) + {\mathcal O}(\alpha_s^2). 
\end{equation}  
We have then
\begin{eqnarray} 
   Z_3 Z_G = 1-\frac{\alpha_s}{4\pi} \left ( \frac{2}{\epsilon} -\gamma + \ln 4\pi \right) \biggr [ 
       -\frac{11}{3} N_c +\frac{2}{3} n_f \biggr ] + {\mathcal O}(\alpha_s^2),        
\end{eqnarray} 
Therefore, the $\mu$-evolution of quasi gluon GPDs is
\begin{equation} 
 \frac{d}{d\ln\mu^2} {\mathcal F}_g^{\mu\nu} (z, \xi, t) = -\frac{\alpha_s \beta_0 }{4\pi}  {\mathcal F}_g^{\mu\nu} (z, \xi, t) +{\mathcal O}(\alpha_s^2),   
 \end{equation} 
 with 
 \begin{equation} 
   \beta_0 =-\frac{11}{3} N_c +\frac{2}{3} n_f.  
\end{equation}    
From results in Sect. 3 there is no mixing with quark quasi GPDs.  
It is noted that the constant $Z_3 Z_G$ is the gluon wave function renormalization constant in  the axial gauge $ n_z \cdot G =0$. In this gauge we have:
  \begin{equation} 
     Z_g \sqrt{Z_3 Z_G} =1
\end{equation} 
where $Z_g$ is the renormalization constant of the coupling.       
This result implies that the product 
$g_s G^\mu$  does not need to be renormalized in the axial gauge as expected.

\par\vskip10pt
\begin{figure}[hbt]
\begin{center}
\includegraphics[width=10cm]{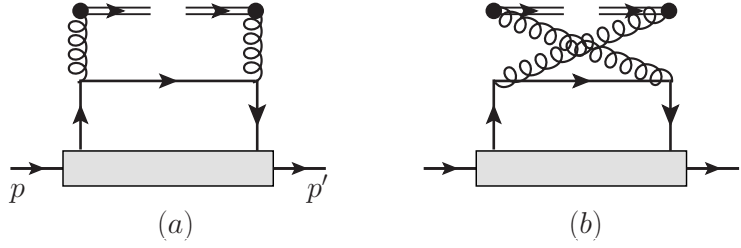}
\end{center}
\caption{The one-loop diagram for quasi GPDs. It represents contributions from quark GPDs.      } 
\label{Fig3}
\end{figure}

\noindent 
{\bf 3.3. One-Loop Contribution from quark GPDs} 

\par\vskip5pt

The quark GPDs will also contribute to the quasi gluon GPDs. The relevant twist-2 quark GPDs are defined and parameterized for a spin-1/2 hadron as: 
\begin{eqnarray}
 F_q (x,\xi, t) &=& \frac{1}{2}  \int \frac{d\lambda}{2\pi} 
                e ^{i x  P^+ \lambda }
                \langle p'  \vert
                \bar\psi (-\frac{\lambda}{2} n )  {\mathcal L}_n^\dagger (-\frac{\lambda}{2},\infty) \gamma^+ {\mathcal L}_n  (\frac{\lambda}{2},\infty)
                  \psi(\frac{\lambda}{2}  n)\vert p \rangle 
\nonumber\\
         &=& \frac{1}{2 P^+} \bar u(p') \biggr [ \gamma^+ H_q (x,\xi, t) + i \frac{ \sigma^{+\nu} \Delta_\nu}{2 m} 
            E_q (x,\xi,t) \biggr ] u(p),  
\nonumber\\
   F_{qL}  (x,\xi, t) &=& \frac{1}{2}  \int \frac{d\lambda}{2\pi} 
                e ^{i x  P^+ \lambda }
                \langle p'  \vert
                \bar\psi (-\frac{\lambda}{2} n )  {\mathcal L}_n^\dagger (-\frac{\lambda}{2},\infty) \gamma^+ \gamma_5 {\mathcal L}_n  (\frac{\lambda}{2},\infty)
                  \psi(\frac{\lambda}{2}  n)\vert p \rangle
\nonumber\\
         &=& \frac{1}{2 P^+} \bar u(p') \biggr [ \gamma^+\gamma_5   H_{qL} (x,\xi, t) +  \frac{ \gamma_5 \Delta^+   }{2 m} 
             E_{qL} (x,\xi,t) \biggr ] u(p)   
\label{DGPD}                                                    
\end{eqnarray}
where the gauge links are defined in the fundamental representation of $SU(N_c)$.  These quark GPDs are for unpolarized- and longitudinally polarized quark, respectively.   
At twist-2 there are quark GPDs for transversely polarized quarks. But it gives no contribution here because of 
the helicity conservation.

The contribution to quasi gluon GPDs comes from diagrams in Fig.5. 
The leading power contribution from Fig.5a  after working out color- and other trivial factors is
\begin{eqnarray} 
 {\mathcal F}_g^{\mu\nu}  (z,\xi,t)\biggr\vert_{5a}  &=& i g_s^2 C_F \frac{P^+}{P_z} \int d x \int \frac{d^4k}{(2\pi)^4} \frac{\delta ( (\hat k_1+\hat k_2)^z/2 - k^z - z P_z) 
} {((\hat k_1 -k)^2+i\varepsilon)((\hat k_2 -k)^2+i\varepsilon) ( k^2 + i\varepsilon)} 
\nonumber\\
&& ((\hat k_1-k)\cdot n_z g^{\mu\sigma} -(\hat k_1-k)^\mu n_z^\sigma)  ((\hat k_2-k)\cdot n_z g^{\nu\rho} -(\hat k_2-k)^\nu n_z^\rho) 
\nonumber\\
    && \biggr [ \gamma_\rho\gamma\cdot k \gamma_\sigma  \biggr ]_{ij} 
\int \frac{d \lambda} {2\pi} e^{i x P^+ \lambda} \langle p' \vert \bar\psi_i  (-\lambda n/2) \psi_j  (\lambda n /2) \vert p\rangle,     
\end{eqnarray}   
where $ij$ are Dirac indices. We only need to consider the case where $\mu$ and $\nu$ are transverse. The quark density matrix is related to quark GPDs:
\begin{equation} 
\int \frac{d \lambda} {2\pi} e^{i x P^+ \lambda} \langle p' \vert \bar\psi_i  (-\lambda n/2) \psi_j  (\lambda n /2) \vert p\rangle =\frac{1}{2} \gamma^- F_q (x,\xi,t) +\frac{1}{2} \gamma_5 \gamma^- F_{qL} (x,\xi,t) 
  +\cdots, 
\end{equation}   
where $\cdots$ are for irrelevant terms.   
Performing the loop integration of the momentum $k$, we have the result: 
\begin{eqnarray} 
 {\mathcal F}_g^{\mu\nu}  (z,\xi,t)\biggr\vert_{5a}  &=& \frac{\alpha_s}{8\pi}  C_F   \int d x
 \biggr \{   g_\perp^{\mu\nu} F_q (x,\xi,t) \biggr [   \frac{\vert z+\xi \vert }{2\xi (x+\xi)} (2x+\xi -z)  \biggr (\frac{2}{\epsilon_c} 
    +\ln\frac{4\pi \mu_c^2}{ (2 P_z (z+\xi))^2 e^\gamma}\biggr )   
\nonumber\\
      && + \frac{ (z-x)^2-\xi^2 + x^2}{2(x^2-\xi^2) } \epsilon(x-z) \biggr (\frac{2}{\epsilon_c} 
    +\ln\frac{4\pi \mu_c^2}{ (2 P_z (x-z))^2 e^\gamma}\biggr )       
\nonumber\\    
   && +\frac{z+\xi}{2\xi (x+\xi)} (4x-z-\xi)\epsilon(z+\xi) + \frac{x-z}{2(x^2-\xi^2)} (5x-z) \epsilon(x-z)
\nonumber\\    
   && +(\xi\to -\xi) \biggr ] +     
    i \epsilon_\perp^{\mu\nu}  F_{qL}  (x,\xi,t)   
 \biggr [  \frac{ (z+\xi)^2} { 2\xi (x +\xi)  } \epsilon(z+\xi) \biggr (\frac{2}{\epsilon_c} 
    +\ln\frac{4\pi \mu_c^2}{ (2 P_z (z+\xi))^2 e^\gamma}\biggr ) 
\nonumber\\
    && - \frac{ z^2+\xi^2-2 x z}{2 (x^2-\xi^2)} \epsilon(x-z) \biggr ( \frac{2}{\epsilon_c} + \ln \frac{4\pi \mu_c^2}{e^\gamma (2P_z (x-z) )^2} \biggr ) 
\nonumber\\    
    &&  + 2z \frac{\vert z+\xi\vert}{\xi (x+\xi)} + 2z \frac{\vert x-z\vert }{x^2-\xi^2} +(\xi\to -\xi) \biggr ]  \biggr \} . \end{eqnarray}             
The contribution from Fig.5b can be obtained from that of Fig.5a as:
\begin{equation} 
   {\mathcal F}_g^{\mu\nu}  (z,\xi,t)\biggr\vert_{5b} ={\mathcal F}_g^{\nu\mu}  (-z,\xi,t)\biggr\vert_{5a}. 
\end{equation}    
There is no U.V. divergence in the quark contribution at the order.  

Again there is the double counting of contributions from collinear regions of loop momentum as discussed before. A subtraction is needed to obtain the correct result. For this we need to calculate the same diagrams in Fig.5 in which the double lines represent the gauge links along the light-cone $n$-direction. With the obtained results, the subtraction is obtained with the same replacement as given in Eq.(\ref{SUBC}).

 \par\vskip20pt
 \noindent
 {\bf 4. Complete Results and the Forward Limit} 
 
 \par\vskip5pt
In this section we give our complete results for the factorization relation of gluon quasi GPDs, which are factorized with twist-2 GPDs. The factorization relation for the three quasi GPDs defined in Eq.(\ref{QGGPD}) can be written in the form:  
\begin{eqnarray}
\mathcal {F}_{gU} (z,\xi,t) &=& \int_{-1}^1 dx \Big[
H_{Ugg}(z,x,\xi) F_{gU}(x,\xi,t)
+H_{Ugq}(z,x,\xi) F_q(x,\xi,t)\Big],
\nonumber\\
\mathcal {F}_{gL}(z,\xi,t) &=& \int_{-1}^1 dx \Big[
H_{Lgg}(z,x,\xi) F_{gL}(x,\xi,t)
+H_{Lgq}(z,x,\xi) F_{qL}(x,\xi,t)\Big],
\nonumber\\
\mathcal {F}^{\mu\nu}_{gT}(z,\xi,t) &=& \int_{-1}^1 dx 
H_{Tgg}(z,x,\xi) F_{gT}^{\mu\nu}(x,\xi,t), 
\label{FACG}
\end{eqnarray}
where all $H$'s are perturbative coefficient functions. 

The perturbative coefficient functions for unpolarized quasi gluon GPDs are:
\begin{eqnarray}
H_{Ugq} (z,x,\xi) &=& -\frac{\alpha_s C_F}{4\pi}
\Big [ \epsilon(x-z)\Big ( \frac{\xi^2 -x^2 -(x-z)^2}{
   x^2-\xi
   ^2}
   \ln\frac{\mu^2}{(2 (x-z)P_z)^2}
      -\frac{3 x^2-4 x
   z+\xi ^2}{x^2-\xi ^2}\Big )   
\nonumber\\   
   && -\epsilon(z-\xi)\frac{ x (z-\xi
   )^2}{ \xi
   \left(x^2-\xi ^2\right)}\ln\frac{\mu^2}{(2(z-\xi)P_z)^2}
   +\frac{4 x |z-\xi|}{x^2-\xi^2} +(z\rightarrow -z)\Big ]  ,
\nonumber\\
H_{Ugg} (z,x,\xi)
 &=& \delta(x-z)-\frac{\alpha_s N_c}{4\pi}
\Big [ 
\delta(x-z)V_0 (z,x,\xi) +V_{U0} (z,x,\xi)
+\ln\frac{\mu^2}{(2P_z)^2}
\Big (  V_{UL}  (z,x,\xi) 
\nonumber\\
  && -\delta(x-z)\ln\frac{|x^2-\xi^2|}{1-x^2}   \Big) +(z\rightarrow -z)\Big ] ,
\end{eqnarray}
with $V_{0,U0,UL}$ given in the below:
\begin{eqnarray}
 V_0  (z,x,\xi)  &=& 1-\frac{\pi^2}{3}-\ln\frac{|x^2-\xi^2|}{1-x^2}+\ln^2|x-\xi|+\ln^2|x+\xi|
-\ln^2(1-x)-\ln^2(1+x),  
\nonumber\\
V_{UL}(z,x,\xi) &=&  \epsilon(z-\xi)\frac{(z-\xi )^2 \left(x^4+x^2 \left(\xi
   ^2-z^2\right)+z \xi  \left(2 z^2+3 z \xi +2
   \xi ^2\right)\right)}{2\xi\left(x^2-z^2\right) (x^2-\xi^2)^2}
\nonumber\\
  && +\frac{ |x-z| \left(x^2+z^2\right)}{(x^2-\xi^2)^2}
+\frac{\left(z^2-\xi ^2\right)}{(x^2-\xi^2)|x-z|_+}, 
 \nonumber\\
V_{U0} (z,x,\xi)  &=& \frac{z^2-\xi^2}{(x^2-\xi^2)|x-z|_+}
-2\frac{(z^2-\xi^2)}{x^2-\xi^2}\Big(\frac{\ln |x-z|}{|x-z|}\Big)_+
-\frac{2|x-z|(x^2+z^2)}{(x^2-\xi^2)^2}\ln|x-z| 
\nonumber\\
 &&+\frac{|x-z|}{3(x^2-\xi^2)^2}\Big[5x^2-xz+5z^2+9\xi^2\Big] -  \frac{\epsilon(z-\xi)(z-\xi)^2\ln|z-\xi|}{\xi(x^2-z^2)(x^2-\xi^2)^2}
\Big[x^4
 \nonumber\\ 
     && +x^2(\xi^2-z^2) +z\xi(2z^2+3z\xi+2\xi^2)\Big] -\frac{|z-\xi|}{3(x^2-z^2)(x^2-\xi^2)^2}
\Big[
6x^4 
\nonumber\\
 &&  -4x^2(z-\xi)(z+2\xi) +z(-5z^3+z^2\xi-5z\xi^2+3\xi^3)
\Big]. 
\label{eq:Vi} 
\end{eqnarray}
The function $V_0$ also appears in other two factorization relations in Eq.(\ref{FACG}). 
The perturbative coefficient functions for longitudinally polarized quasi gluon GPDs are:
\begin{eqnarray}
H_{Lgq} (z,x,\xi)  &=& \frac{\alpha_s C_F}{4\pi}
\Big[\epsilon(x-z)\frac{2 x z-z^2-\xi^2}{
   \left(x^2-\xi
   ^2\right)}\ln\frac{\mu^2}{
   (2 (x-z)P_z)^2   }
   +\frac{4 z
   |x-z|}{x^2-\xi ^2}
\nonumber\\   
  &&  +\epsilon(z-\xi)\frac{ (z-\xi )^2}{
   \left(\xi
   ^2-x^2\right)}\ln\frac{\mu^2}{
   (2(z-\xi)P_z)^2   }
   +\frac{4 z
   |z-\xi|}{\xi ^2-x^2}-(z\rightarrow -z)\Big],
\nonumber\\
H_{Lgg} (z,x,\xi) 
&=& \delta(x-z)-\frac{\alpha_s N_c}{4\pi}
\Big\{
\delta(x-z)V_0 (z,x,\xi) +V_{L0}(z,x,\xi)
\nonumber\\
 &&+\ln\frac{\mu^2}{(2P_z)^2}
\Big[-\delta(x-z)\ln\frac{|x^2-\xi^2|}{1-x^2}
+V_{LL} (z,x,\xi)  \Big]
-(z\rightarrow -z)\Big\}
\end{eqnarray}
with $V_{L0,LL}$ as functions depending on $x$, $z$ and $\xi$: 
\begin{eqnarray}
V_{LL} (z,x,\xi) & =& \frac{2 x z |x-z|}{(x^2-\xi^2)^2}+\frac{
   \left(z^2-\xi ^2\right)}{(x^2-\xi^2)|x-z|_+} -\epsilon(z-\xi)\frac{ x (z-\xi )^2 \left(x^2-2 z^2-2 z \xi
   -\xi ^2\right)}{\left(x^2-z^2\right)
   \left(x^2-\xi ^2\right)^2},
\nonumber\\
V_{L0} (z,x,\xi)  &=& -\frac{|x-z| \left(4 x z \ln (|x-z|)-5 x
   z-\xi ^2\right)}{(x^2-\xi^2)^2}-2\Big(\frac{\ln|x-z|}{|x-z|}\Big)_+
   \frac{z^2-\xi ^2}{x^2-\xi^2}
\nonumber\\
  &&+\frac{z^2-\xi
   ^2}{(x^2-\xi^2)|x-z|_+} -\frac{x |z-\xi| }{\left(x^2-z^2\right)
   \left(x^2-\xi ^2\right)^2}
   \Big(4 x^2 z-5 z^3
\nonumber\\   
   &&-2 \left(x^2 (z-\xi
   )-2 z^3+z \xi ^2+\xi ^3\right) \ln (|z-\xi
   |) -z^2 \xi +z \xi ^2+\xi
   ^3\Big) .
\end{eqnarray}
The perturbative coefficient function for transversely polarized quasi gluon GPDs is:
\begin{eqnarray}
H_{Tgg} (z,x,\xi) &=& \delta (x-z) -\frac{\alpha_s N_c}{4\pi}
\Big\{
\delta(x-z)V_0 (z,x,\xi) +V_{T0}(z,x,\xi)
\nonumber\\
 &&+\ln\frac{\mu^2}{(2P_z)^2}
\Big[-\delta(x-z)\ln\frac{|x^2-\xi^2|}{1-x^2}
+V_{TL}(z,x,\xi) \Big]
+(z\rightarrow -z)\Big\}, 
\end{eqnarray}
with 
\begin{eqnarray}
V_{TL} (z,x,\xi) &=& \frac{2\xi^2 |x-z|}{(x^2-\xi^2)^2}+\frac{
   \left(z^2-\xi ^2\right)}{(x^2-\xi^2)|x-z|_+}
 + \epsilon(z-\xi)\frac{(z-\xi )^2 \left(x^2 (z+2 \xi )+z \xi
   ^2\right)}{\left(x^2-z^2\right)
   \left(x^2-\xi^2\right)^2},
\nonumber\\
V_{T0} (z,x,\xi)  &=& \frac{|x-z| \left(-12 \xi ^2
   \ln|x-z|+x^2+x z+z^2+15 \xi
   ^2\right)}{3(x^2-\xi^2)^2}-2\Big(\frac{\ln |x-z|}{|x-z|}\Big)_+
  \frac{(z^2-\xi ^2)}{x^2-\xi^2}
\nonumber\\
  &&+\frac{
   \left(z^2-\xi ^2\right)}{(x^2-\xi^2)|x-z|_+} + \frac{|z-\xi| }{3 \left(x^2-z^2\right)
   \left(x^2-\xi ^2\right)^2}
   \Big(2 x^2 \left(z^2+z \xi -8 \xi ^2\right) 
 \nonumber\\
   &&-6 (z-\xi ) \left(x^2
   (z+2 \xi )+z \xi ^2\right) \ln (|z-\xi
   |)+z
   \left(z^3+z^2 \xi +13 z \xi ^2-3 \xi
   ^3\right)\Big).
\end{eqnarray}

With the above results in Eq.(\ref{FACG}) and the renormalization of quasi gluon GPDs discussed at the end of the subsection 3.2., we can derive the $\mu$-evolution equations of gluon GPDs. The results are:
\begin{eqnarray} 
  \frac{d}{d\ln \mu^2} F_{gU} (z,\xi,t) &=& \frac{\alpha_s}{4\pi} \int_{-1}^1 dx \Big[
  {\mathcal K}_{Ugg}(z,x,\xi) F_{gU}(x,\xi,t)
+{\mathcal K}_{Ugq}(z,x,\xi) F_q(x,\xi,t)\Big],
\nonumber\\
\frac{d}{d\ln \mu^2}F_{gL}(z,\xi,t) &=&  \frac{\alpha_s}{4\pi} \int_{-1}^1 dx \Big[
{\mathcal K}_{Lgg}(z,x,\xi) F_{gL}(x,\xi,t)
+{\mathcal K}_{Lgq}(z,x,\xi) F_{qL}(x,\xi,t)\Big],
\nonumber\\
\frac{d}{d\ln \mu^2} F^{\mu\nu}_{gT}(z,\xi,t) &=& \frac{\alpha_s}{4\pi}  \int_{-1}^1 dx 
{\mathcal K} _{Tgg}(z,x,\xi) F_{gT}^{\mu\nu}(x,\xi,t), 
\end{eqnarray} 
with ${\mathcal K}$'s as evolution kernels given in the below: 
\begin{eqnarray} 
 {\mathcal K}_{Agg}(z,x,\xi) &=&  \delta (x-z)  \Big ( -\frac{\beta_0}{2}  - N_c \ln \frac{\vert x^2-\xi^2\vert}{1-x^2} \Big ) 
      + N_c V_{AL} (z,x,\xi) + (z\to -z),  
 \nonumber\\
 {\mathcal K}_{Ugq}(z,x,\xi) &=& C_F
\Big [ \epsilon(x-z) \frac{\left(-2 x^2+2 x
   z-z^2+\xi
   ^2\right)}{
   \left(x^2-\xi^2\right)} 
  -\epsilon(z-\xi)\frac{ x (z-\xi
   )^2}{ \xi
   \left(x^2-\xi ^2\right)} +(z\rightarrow -z)\Big ]  ,
\nonumber\\
{\mathcal K}_{Lgq} (z,x,\xi)  &=& - C_F 
\Big[\epsilon(x-z)\frac{2 x z-z^2-\xi^2}{
   \left(x^2-\xi
   ^2\right) } 
   +\epsilon(z-\xi)\frac{ (z-\xi )^2}{
   \left(\xi
   ^2-x^2\right)}-(z\rightarrow -z)\Big],
\nonumber\\
{\mathcal K}_{Lgg} (z,x,\xi) 
 &=& \delta (x-z) \Big (- \frac{\beta_0}{2} -N_c \ln \frac{\vert x^2-\xi^2\vert}{1-x^2} \Big )  + N_c V_{LL} (z,x,\xi)  
-(z\rightarrow -z), 
\end{eqnarray} 
where $A$ in the first line stands for $U$ or $T$.  These evolution equations have been studied in \cite{Ra,BGR}   
and summarized in \cite{BeRa}. Our derived evolution equations agree.  From our results of the factorization,  gluon quasi GPDs depend on $P_z$  through $\ln P_z$ at order of $\alpha_s$ if one neglects higher-twist effects.   
This dependence can be read off from the perturbative coefficient functions in Eq.(\ref{FACG}).  

\par
In the forward limit, i.e., the limit with $\Delta^\mu\to 0$, only two GPDs of a spin-1/2 hadron in Eq.(\ref{GT1}) survive. Similarly, two quark GPDs in Eq.(\ref{DGPD}) are nonzero in the limit. These GPDs are related to the standard twist-2 PDFs as:
\begin{equation} 
 H_g (x,0,0) = x f_g (x), \quad H_{gL} = x f_{gL} (x), \quad H_q (x,0,0) = f_q (x), \quad 
  H_{qL} (x,0,0) = f_{qL}(x),  
\end{equation}   
where $f_g$ and $f_{gL}$ are twist-2 unpolarized- and  longitudinally polarized gluon PDF, respectively. 
$f_{q, qL}$ are the corresponding quark PDFs. In the forward limit, quasi gluon GPDs become quasi gluon PDFs. 
There are two quasi gluon PDFs from Eq.(\ref{GT2}) in the limit:
\begin{equation}  
 {\mathcal H}_g (z,0,0) = z \tilde f_g (z), \quad {\mathcal H}_{gL}(z,0,0) = z \tilde f_{gL} (z),   
 \end{equation} 
with $\tilde f_{g,gL}$ are corresponding quasi gluon PDFs. 
Taking the forward limit in the results for quasi gluon GPDs, we obtain the factorization relation 
of quasi gluon PDFs with twist-2 PDFs:
\begin{eqnarray}
\tilde{f}_g(z) &=& \int_{-1}^1 dx C_{Ugg}(z,x) f_g(x)
+\int_{-1}^1 dx C_{Ugq}(z,x)f_q(x),
\nonumber\\
\tilde{f}_{gL}(z) &=& \int_{-1}^1 dx C_{Lgg}(z,x) f_{gL}(x)
+\int_{-1}^1 dx C_{Lgq}(z,x)f_{qL}(x). 
\end{eqnarray}
The perturbative coefficient functions of the unpolarized quasi gluon PDF are:  
\begin{eqnarray}
C_{Ugq}(z,x)
 &=& \frac{\alpha_s C_F}{4\pi}\frac{1}{z}\Big\{
\ln\frac{\mu^2}{(2P_z)^2}
\Big[\epsilon(x-z)\frac{-2x^2+2xz-z^2}{x^2}+\epsilon(z)\frac{2z}{x}\Big] +\epsilon(z)\frac{2z}{x}(1-\ln z^2)
\nonumber\\
 &&+\epsilon(x-z)\Big(
\frac{2x^2-2xz+z^2}{x^2}\ln(x-z)^2-3+\frac{4z}{x}
\Big) + (z\rightarrow -z)
\Big\}, 
\nonumber\\
C_{Ugg}(z,x) &=& \delta (x-z)-\frac{\alpha_s N_c}{4\pi}
\Big\{
\delta(x-z)\tilde{V}_0(z,x)+\tilde V_{U0}(z,x)
\nonumber\\
&&+\ln\frac{\mu^2}{(2P_z)^2}\Big[-\delta(x-z)\ln\frac{x^2}{1-x^2}+\tilde V_{UL}(z,x)\Big]- (z\rightarrow -z)\Big\},
\label{eq:cUgg}
\end{eqnarray}
where
\begin{eqnarray}
\tilde{V}_0(z,x) &=& 1-\frac{\pi^2}{3}-\ln\frac{x^2}{1-x^2}+\frac{1}{2}\ln^2x^2
-\ln^2(1-x)-\ln^2(1+x),
\nonumber\\
\tilde{V}_{UL} (z,x) &=& \frac{(x^2-xz+z^2)^2}{zx^3|x-z|_+}
   +\frac{\epsilon(z)}{x^3(x^2-z^2)}[-x^4+x^2 z^2 +z^4],
\nonumber\\
\tilde{V}_{U0}(z,x) & =& \frac{z}{x|x-z|_+}
-2\frac{z}{x}\Big(\frac{\ln|x-z|}{|x-z|}\Big)_+
-\frac{|x-z|}{3x^3 z}\Big[-5x^2+xz-5z^2+3(x^2+z^2)\ln(x-z)^2\Big]
\nonumber\\
 && + \frac{\epsilon(z)}{3x^3(x^2-z^2)}\Big[
-3x^4+x^2 z^2 +5z^4 +3(x^4-x^2 z^2-z^4)\ln z^2\Big].
\end{eqnarray}
The perturbative coefficient functions of the  longitudinally polarized quasi gluon PDF are:
\begin{eqnarray}
C_{Lgq}(z,x)
 &=& -\frac{\alpha_s C_F }{4\pi x^2}\Big\{
  \ln\frac{\mu^2}{(2P_z)^2}
\Big[\epsilon(x-z)(2x-z)- z \epsilon(z)\Big]+\epsilon(z) z (\ln z^2 -4)
\nonumber\\
 &&+\epsilon(x-z)  \Big[(z-2x)\ln(x-z)^2+4x-4z\Big]
    + (z\rightarrow -z)
\Big\}, 
\nonumber\\
C_{Lgg}(z,x) &=&\delta (x-z) -\frac{\alpha_s N_c}{4\pi}
\Big\{
\delta(x-z)\tilde{V}_0+\tilde{V}_{L0}+ \ln\frac{\mu^2}{(2P_z)^2}\Big[-\delta(x-z)\ln\frac{x^2}{1-x^2}+\tilde{V}_{LL}\Big]
\nonumber\\
 &&+(z\rightarrow -z)\Big\},
\end{eqnarray}
with
\begin{eqnarray}
\tilde{V}_{LL}  &=& \frac{z}{x|x-z|_+}+\frac{2|x-z|}{x^2} +z\epsilon(z)\frac{2z^2 -x^2}{x^2 (x^2-z^2)},
\nonumber\\
\tilde{V}_{L0} &=& \frac{z}{x|x-z|_+}
-2\frac{z}{x}\Big(\frac{\ln |x-z|}{|x-z|}\Big)_+
+\frac{5|x-z|}{x^2}-\frac{2|x-z|}{x^2}\ln (x-z)^2,
\nonumber\\
&&  +\frac{|z|}{x^2(x^2-z^2)}\Big[
-4x^2+5z^2+(x^2-2z^2)\ln z^2\Big].
\end{eqnarray}
The factorization of quasi gluon PDFs has been studied in \cite{WZZ}. Our one-loop result 
agrees with that in  \cite{WZZ}, if the flavors of anti-quarks are not only included in the integration but also in the sum in Eq.(70) of \cite{WZZ},  and $P^z$ is replaced with $yP^z$ in the convolution kernal. 
Without the replacement DGLAP equation of gluon PDF can not be derived correctly. It has been  also pointed out in \cite{IJJSZ} that in such a convolution as Eq.(70) of \cite{WZZ} the correct variable is $yP^z$ instead of $P^z$.

Before summarizing our work,  we note that the perturbative coefficient functions are proportional to $z^2$ or $\vert z\vert $ for $\vert z\vert \to \infty$. This implies that moments of gluon quasi GPDs cannot defined, i.e., integrals like 
\begin{equation} 
    \int_{-\infty}^{\infty} d z z^n {\mathcal F}^{\mu\nu} (z,\xi,t) 
\end{equation} 
are divergent at least for positive $n$. This fact prevents from the expansion of the nonlocal operators 
for quasi GPDs into local operators.

\par\vskip40pt
\noindent 
{\bf 5. Summary} 
\par\vskip5pt
In this work, 
we have shown that at one-loop level quasi gluon GPDs can be factorized with twist-2 GPDs in the limit 
of large hadron momentum. The perturbative coefficient functions in the factorization are free from any collinear- or I.R.  divergences. This implies that quasi gluon GPDs contain the same long-distance effects as twist-2 GPDs in the limit. In the derivation of our results, we have to include ghost contributions in order to keep the gauge-invariance of the factorization. Our work shows that the operator-mixing pattern 
of nonlocal operators used to define quasi gluon GPDs is the same as that  proven for local operators, i.e., 
they are only mixed with gauge-invariant operators, BRST-variation operators and operators involving EOM operator. 
Our results are complete for all  quasi gluon GPDs and can be used not only 
for hadrons with 1/2-spin but also for hadrons with spins other than 1/2.  Taking the forward limit we obtain the factorization relation between 
quasi gluon PDFs and twist-2 PDFs from our results.        

\par\vskip40pt

\noindent
{\bf Acknowledgments}
\par
The work is supported by National Natural Science Foundation of P.R. China(No.12075299,11821505, 11847612,11935017 and 12065024)  and by the Strategic Priority Research Program of Chinese Academy of Sciences, Grant No. XDB34000000.

\par

\end{document}